# Kinematic wave solutions for dam-break floods in non-uniform valleys

Tzu-Yin Kasha Chen[a] and Hervé Capart[b]

[ab]Department of Civil Engineering and Hydrotech Research Institute,

National Taiwan University, Taipei 106, Taiwan

Email: [a] D06521004@ntu.edu.tw, [b] hcapart@yahoo.com

**Abstract**

In non-uniform valleys, dam-break flood waves can be significantly affected by downstream variations in river width, slope and roughness. To model these effects, we derive new analytical solutions to the kinematic wave equation, applicable to rating curves in the power law form and hydrographs of generic shape as long as they produce a single shock at the wave front. New results are first obtained for uniform channels, using the Gauss-Green theorem applied to characteristic-bounded regions of the ($x,t$) plane. The results are then extended to non-uniform valleys, using a change of variable that homogenizes river properties by rescaling the distance coordinate. The solutions are illustrated and validated for idealized cases, then applied to three historical dam failures: the 2008 breaching failure of the Tangjiashan landslide dam, the 1976 piping failure of Teton Dam, and the 1959 sudden failure of Malpasset Dam. In spite of the much reduced computational cost and data requirements, the results agree well with the field data and with more elaborate simulations. They also clarify how both river and hydrograph properties affect flood propagation and attenuation.

**Key words:** kinematic wave; dam break flood; flood propagation; non-uniform valley.

# 1. Introduction

The simulation of dam failure floods is an important tool to evaluate risk exposure and draw up evacuation plans downstream of dams. After an upstream discharge hydrograph has been estimated based on a dam failure scenario, the resulting flood can be simulated using a range of different methods. Currently, the most elaborate approach is to simulate flood propagation using detailed dynamic models based on the two-dimensional shallow water equations (Valiani et al., 2002; Carrivick, 2006; Fan et al., 2012; Kim and Sanders, 2016; Stilmant et al., 2018). Possibly the simplest approach, by contrast, is to assume a uniform valley and model the flood using the one-dimensional kinematic wave equation (Lighthill and Whitham, 1955; Whitham,1974; Hunt, 1984a, b, 1995; Kazezyilmaz-Alhan and Medina, 2007; Capart, 2013; Stilmant et al., 2018).

For end-users, it is necessary to use two-dimensional dynamic models whenever high-resolution results are important. Such models are for instance needed to obtain detailed flood depths and scour velocities, or to estimate impact forces on structures (Aureli et al., 2015). Because they require large computation times and detailed valley topography data, on the other hand, such models may become impractical when a rapid assessment is needed. According to FEMA (2013), dam breach formation time is typically short, e.g. 0.1 to 4 hours for earth-fill dams, 0.1 to 0.5 hours for concrete gravity dams, and smaller than 0.1 hours for concrete arch dams. For emergency decisions regarding downstream evacuation, therefore, simpler models are needed. As they require less computational efforts and simpler cross-section data, one-dimensional kinematic models thus become an attractive option. For these models, it is possible to derive analytical solutions for various special cases (Hunt 1984a, b; Capart, 2013) and efficient numerical solutions for more general problems (Hunt, 1995; Kazezyilmaz-Alhan and Medina, 2007; Chen, 2017). In spite of their simplicity, kinematic wave models have been found to yield accurate results for floods that develop over long distances, and when the valley or channel is nearly uniform (Bohorquez, 2010; Capart 2013; Stilmant et al., 2018).

In practice, unfortunately, valleys downstream of dams typically feature large variations in properties like river width, slope and roughness. Both engineered and natural dams are often located in narrow, steep upstream valleys, whereas the exposed populated areas lie in wider, less inclined downstream valleys or plains. An example in Sichuan Province, China, is illustrated in Fig. 1, where a landslide dam formed and breached following the 2008 Wenchuan earthquake (Lui et al., 2009). This landslide dam blocked Tongkou River and formed a barrier lake in the upstream narrow valley. Downstream, the valley first narrows as it crosses a fault belt, then widens in the plain area. In the present paper, we examine whether the kinematic wave model could take such variations into consideration, yet still allow the derivation of simple analytical solutions.

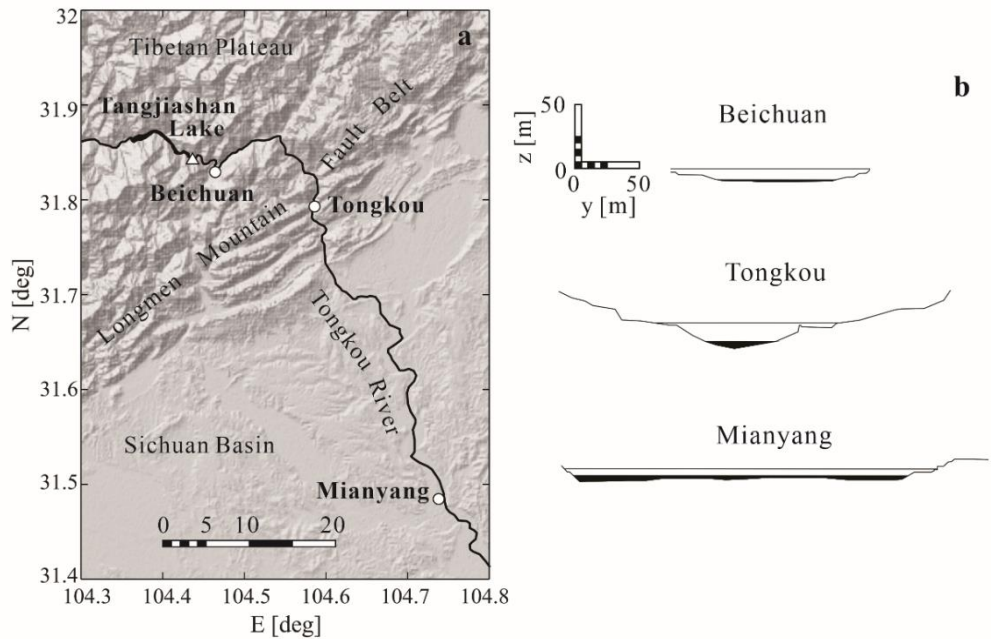

**Fig. 1.** Site of the Tangjiashan landslide dam (after Capart, 2013, based on data from GMTED2010, USGS Global Multi-resolution Terrain Elevation Data, 2010). (a) Shaded relief view of the regional topography with locations of the landslide dam (triangle) and flood gauging stations along Tongkou River (circles); (b) river cross-sections at the Beichuan, Tongkou, and Mianyang gauging stations (data source: Liu et al., 2009).

We find that this is indeed possible, subject to some restrictions: the upstream discharge hydrograph must produce a single shock at the wave front, and the dependences of the rating curve on discharge and distance must be separable. This last requirement is satisfied for rating curves in power law form, which is a common modeling choice (Hunt, 1984a; Kazezyilmaz-Alhan and Medina, 2007; Capart, 2013). The exponent of the rating curves must be set to describe either U-shaped or V-shaped cross sections, but the coefficient of the rating curve can vary freely with downstream distance. Note that, for both uniform and non-uniform valleys, kinematic models cannot account for non-local backwater and inertial effects. Therefore, they cannot be applied to adverse slope reaches or to the lake upstream of a dam (Hunt, 1995; Singh, 2001).

The paper is structured as follows. In section 2, we describe the equations and assumptions of our kinematic wave model, and explain how solution properties like characteristic paths are affected by variations in valley properties. In section 3, we derive new solutions for uniform valleys, needed to simulate flood hydrographs of generic shape. In section 4, we then extend these solutions to non-uniform valleys, using a change of variable that homogenizes river properties by rescaling the distance coordinate. In both these sections, we illustrate and validate these solutions for idealized cases, for which comparisons with previous analytical or numerical results can be made. In section 5, we apply the new model to three well-documented field cases: the 2008 breaching failure of the Tangjiashan landslide dam, the 1976 piping failure of Teton Dam, and the 1959 sudden failure of Malpasset Dam. In all three cases, kinematic model results can be compared with field data, and with dynamic model results by previous researchers (Goutal, 1999; Valiani et al., 2002; Hervouet and Petitjean, 1999; Liu et al., 2009 and 2010; Fan et al., 2012; U.S. Geological Survey, 1976; Gundlach and Thomas, 1977). In section 6, finally, the findings are summarized and conclusions are drawn.

## 2. Assumptions and governing equations

In the kinematic method, one-dimensional flood routing is conducted by solving the continuity equation (Lighthill and Whitham, 1955; Whitham,1974; Hunt, 1984a, 1984b, 1995; Capart, 2013; Stilmant et al., 2018)

$$\frac{\partial A}{\partial t} + \frac{\partial Q}{\partial x} = 0, \tag{1}$$

where $A(x,t)$ is the flow area and $Q(x,t)$ the flow discharge, both expressed in terms of two independent variables, the time $t$ and the distance $x$. Momentum balance is simplified by assuming that the flow is quasi-uniform, and that the downslope pull of gravity is locally equilibrated by friction along the channel perimeter (Lighthill and Whitham, 1955; Whitham,1974; Hunt, 1984a; Kazezyilmaz-Alhan and Medina, 2007; Capart,

2013). As a result, at each cross section the variables $Q$ and $A$ are connected by a one-to-one relation, or rating curve. This can either be calibrated on site, or calculated from an assumed cross section geometry and flow resistance law. To treat non-uniform valleys, we assume a relation of the form

$$A = A(Q, x), \qquad (2)$$

where $Q = Q(x,t)$. In that case, the chain rule can be used to rewrite Eq. (1) in the form

$$\frac{\partial A}{\partial Q}\frac{\partial Q}{\partial t} + \frac{\partial Q}{\partial x} = 0. \qquad (3)$$

Some special properties of the solutions to this equation can be highlighted using the method of characteristics. Consider an observer moving along a path $x_{obs}(t)$. The discharge $Q_{obs}(t) = Q(x_{obs}(t), t)$ observed along this path then varies with time according to the total derivative

$$\frac{dQ_{obs}}{dt} = \frac{\partial Q}{\partial x}\frac{dx_{obs}}{dt} + \frac{\partial Q}{\partial t}. \qquad (4)$$

Substituting Eq. (3) into Eq. (4) yields

$$\frac{dQ_{obs}}{dt} = \left(-\frac{\partial A}{\partial Q}\frac{dx_{obs}}{dt} + 1\right)\frac{\partial Q}{\partial t}. \qquad (5)$$

There then exist special observer paths, known as characteristics, along which

$$\frac{dQ_{obs}}{dt} = 0, \qquad (6)$$

which requires that the observer moves at the characteristic speed

$$\frac{dx_{obs}}{dt} = \lambda(Q_{obs}, x_{obs}) = \frac{1}{\frac{\partial A}{\partial Q}(Q_{obs}, x_{obs})}. \qquad (7)$$

For uniform valleys, the dependence on $x$ disappears, hence the wave speed $\lambda$ is constant along each characteristic. Straight rays are therefore obtained in the $(x,t)$ plane. Since the flow area $A$ depends only on the discharge $Q$, moreover, the flow area like the discharge will be constant along each characteristic. For non-uniform valleys, by contrast, the characteristic speed $\lambda$ varies with $x$, hence the characteristics will be curves in the $(x,t)$ plane. As for light refraction in a non-homogeneous medium, in non-uniform valleys the characteristic rays transporting a given discharge will alter their speeds according to local river properties. Along each characteristic, moreover, the flow area $A_{obs}$ is no longer constant, but varies according to

$$\frac{dA_{obs}}{dt} = \frac{\partial A}{\partial Q}\frac{dQ_{obs}}{dt} + \frac{\partial A}{\partial x}\frac{dx_{obs}}{dt} = \frac{\partial A}{\partial x}\frac{dx_{obs}}{dt}, \tag{8}$$

and only for uniform valleys does this simplify to $dA_{obs}/dt = 0$. Because in general only $Q$ is constant along characteristics, it is simpler to work with rating curves in the form $A = A(Q,x)$ rather than $Q = Q(A,x)$, hence the choice made in Eq. (2). Since the flow area and discharge increase together, note that $\lambda > 0$ hence characteristics always travel downstream.

To simplify, we restrict our attention to rating curves in the power law form (Hunt, 1984a; Kazezyilmaz-Alhan and Medina, 2007; Capart, 2013). These are typically written

$$Q = mA^{\alpha}. \tag{9}$$

We will assume that the coefficient $\alpha$ is constant or piecewise constant, but let the coefficient $m$ vary freely with the distance coordinate $x$. For cases where the coefficient α is piecewise constant, we will apply the constant coefficient solution to each valley segment in turn, taking at each transition the outflow hydrograph of the upstream segment as the boundary condition for the downstream segment. Such piecewise solutions will be described in more detail in section 4.1, then applied to a hypothetical example and a field case in section 4.2 and section 5.2, respectively.

For a single valley segment characterized by a constant coefficient $\alpha$, therefore, the rating curve relation of Eq. (2) will therefore take the form

$$A(Q,x) = \left(\frac{Q}{m(x)}\right)^{1/\alpha}, \tag{10}$$

yielding for the characteristic wave speed of Eq. (7) the result

$$\lambda = \frac{1}{\frac{\partial A}{\partial Q}} = \frac{1}{\frac{1}{\alpha}\left(\frac{Q}{m(x)}\right)^{1/\alpha - 1}\frac{1}{m(x)}} = \alpha m(x)^{1/\alpha} Q^{(\alpha-1)/\alpha}. \tag{11}$$

To determine the coefficient $m(x)$ and the exponent $\alpha$, we assume either a U-shaped or a V-shaped valley, as illustrated in Fig. 2, and apply the Darcy friction formula (Henderson,1966; Hunt, 1984a; Capart, 2013; Stilmant et al., 2018)

$$U = \sqrt{\frac{8gSR}{f}}, \tag{12}$$

where $f$ is the Darcy-Weisbach friction coefficient, $R$ is the hydraulic radius, and $S$ is the local bed slope.

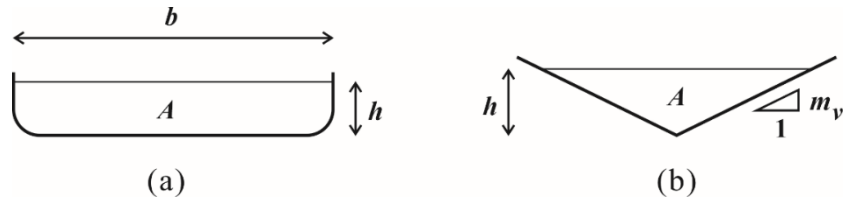

(a)　　　　　　　　　　(b)

**Fig. 2.** Typical valley shapes. (a) U-shape valley; (b) V-shape valley.

For U-shaped reaches, we assume that $h \ll b$, hence $R \cong h$. For valleys of given shape, U or V, the exponent $\alpha$ is constant, but the coefficient $m$ will vary with cross section dimensions (valley width $b$ or side slope $m_v$), and will depend also on the local bed slope $S$ and friction coefficient $f$, which can all vary with distance. The specific formulas can be written

$$\alpha = \frac{3}{2}, \quad m = \sqrt{\frac{8gS}{bf}}, \quad \text{U-shape}, \tag{13}$$

$$\alpha = \frac{5}{4}, \quad m = \sqrt{\frac{4gS}{f}}\left(\frac{m_v}{1+m_v^2}\right)^{1/4}, \quad \text{V-shape}. \tag{14}$$

In this work, we will assume that the valley length can be subdivided into segments characterized by a single shape (U or V), hence piecewise constant values for the exponent $\alpha$. Variations in width, slope, and friction coefficient, on the other hand, will be accommodated by letting the coefficient $m$ vary freely with distance.

Using the kinematic model, flood routing amounts to a signaling problem (Whitham, 1974) that requires initial and upstream boundary conditions. The initial condition is given by the discharge along the valley at $t = 0$. Since the discharge from a dam break flood is usually far larger than typical flows, we adopt the assumption of initially dry bed whereby

$$Q(x,0) = 0. \tag{15}$$

The upstream boundary condition is the discharge hydrograph at the dam assuming some dam failure scenario

$$Q(0,t) = Q_B(t). \tag{16}$$

We consider generic hydrographs, but they must produce only a single shock at the wavefront. No downstream boundary condition is required. Subject to these conditions, we explain in the next section (section 3) how the problem can be solved for the case of a uniform valley ($m = $ constant). The following section (section 4) will then treat the case of non-uniform valleys ($m = m(x)$).

## 3. Solutions for uniform valleys

### 3.1. Solution method

In uniform valleys, $m$ and $\alpha$ are both constant and the wave speed $\lambda$ depends only on the discharge $Q$. The characteristic rays will therefore be straight, because they transport a constant discharge $Q$, but not parallel because the slope of each ray depends on its discharge. Specifically, each ray will travel at the constant speed $\lambda(t_0)=\lambda(Q_B(t_0))$, dependent on the discharge value $Q_B$ at the time $t_0$ when the characteristic first enters the domain at the upstream origin $x=0$. In the $(x,t)$ plane, the characteristic rays thus form a family of straight lines given by

$$x(t;t_0) = \lambda(t_0)(t-t_0), \quad 0 \leq x(t;t_0) \leq x_F(t), \tag{17}$$

where $x_F(t)$ is the time-varying position of the front edge of the flood wave. Along each ray, the discharge $Q$ is equal to $Q_B(t_0)$ and the flow area equal to $A(Q_B(t_0))$. The characteristic rays are therefore also level sets of the discharge and flow area, and the solution is complete provided that the front position $x_F(t)$ can be determined. Starting from an initially dry bed, this front will be discontinuous and its speed governed by the shock condition

$$\frac{dx_F(t)}{dt} = \frac{Q(x_F,t)}{A(Q(x_F,t))}. \tag{18}$$

Various strategies can be adopted to determine the resulting front path $x_F(t)$. The first is to track the shock by integrating in time the ordinary differential Eq. (18), a rather formidable task. The second approach (Hunt, 1984) is to apply conservation of mass in integral form (Gauss-Green theorem) to the rectangular region of the $(x,t)$ plane shown in outline in Fig. 3. For initially dry bed, this yields the equation

$$\int_0^t Q_B(t)dt = \int_0^{x_F(t)} A(x,t)dx, \tag{19}$$

stating that the water volume supplied so far to the valley must be equal to the water volume currently within the valley between 0 and $x_F(t)$. This provides a condition on $x_F(t)$ that can be worked out provided one can derive the distribution of flow area within the valley $A(x,t)$ from Eq. (17). Unfortunately, this will be available in closed form only in exceptional cases.

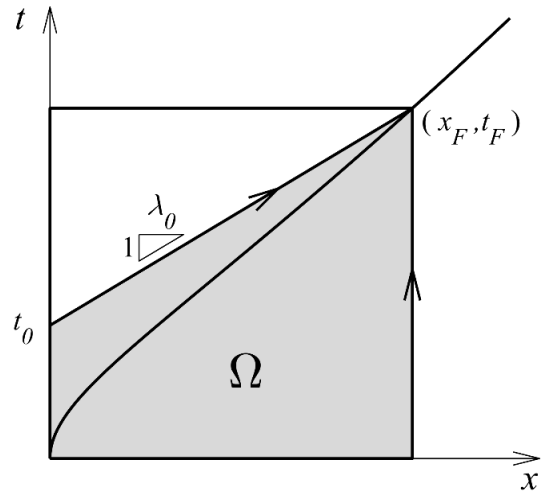

**Fig. 3.** The characteristic ray and the corresponding region $\Omega$ for solving the position of the flood front $(x_F, t_F)$.

To accommodate hydrographs of generic form, we therefore adopt the third approach proposed by Capart (2013, see also Stilmant et al., 2018). In this approach, the Gauss-Green theorem is again used, but this time applied to the characteristics-bounded region shown shaded in Fig. 3. Accordingly, the mass conservation law given by Eq. (1) is integrated over the region $\Omega$ bounded by the characteristic paths that intersect at the shock front. The Gauss-Green theorem is then used to convert the integral over region $\Omega$ to a line integral along its closed boundary $\Gamma$, yielding

$$\iint_\Omega \left( \frac{\partial A}{\partial t} + \frac{\partial Q}{\partial x} \right) = \oint_\Gamma Qdt - Adx = 0. \tag{20}$$

The curve $\Gamma$ can be divided into four straight lines that connect the following vertices: the origin $(0,0)$, the foot $(x_F, 0)$ of the characteristic ray impacting the wave front from downstream (vertical because the valley is initially dry), the point of interest $(x_F, t_F)$ along the wave front, and the point of entry $(0, t_0)$ of the characteristic ray impacting the wave front from upstream. The motivation for choosing this contour over the rectangular contour is that the discharge and flow area are both constant along characteristics. For dry bed initial conditions, only two of these segments contribute to the integral, yielding

$$\int_0^{t_0} Q_B(t)dt + \int_{t_0}^{t_F} Q_B(t_0)dt - \int_0^{x_F(t)} A_B(t_0)dx$$
$$= \int_0^{t_0} Q_B(t)dt + Q_B(t_0)(t_F - t_0) - A_B(t_0)x_F = 0, \quad (21)$$

where $A_B(t_0)$ is the flow area $A(Q_B(t_0))$ associated with $Q_B(t_0)$. Substituting Eq. (10) and (17) into Eq. (21) yields

$$\int_0^{t_0} Q_B(t)dt + (1-\alpha)Q_B(t_0)(t_F - t_0) = 0. \quad (22)$$

Let us define

$$V_B(t_0) = \int_0^{t_0} Q_B(t)dt, \quad (23)$$

representing the total volume of water released to the downstream valley up to time $t_0$. Using Eq. (22), the characteristic originating at point $(0, t_0)$ will then attain the wave front at time $t_F$ given by

$$t_F(t_0) = t_0 + \frac{1}{\alpha}\frac{V_B(t_0)}{Q_B(t_0)}, \quad (24)$$

and the corresponding front position is

$$x_F(t_0) = \frac{\alpha m^{1/\alpha}}{\alpha - 1}\frac{V_B(t_0)}{Q_B(t_0)^{1/\alpha}} = \frac{\alpha}{\alpha - 1}\frac{V_B(t_0)}{A_B(t_0)}. \quad (25)$$

This provides a complete solution for the wave front path $(x_F(t_0), t_F(t_0))$, parameterized according to the time of entry of the characteristic impacting onto the front from upstream. To work out this solution for a generic hydrograph, the only calculation necessary is to integrate the discharge in time using Eq. (23). The solution for the wave front is thus obtained in surprisingly compact form, and applies more generally than the solutions derived earlier for special cases (Hunt, 1984; Capart, 2013). The formulas are of great practical interest because they provide, at very small computational cost, a solution for the downstream time of arrival of the dam-break waves.

The above solution is only valid when the channel bed is initially dry, and when the wave structure includes a single shock at the front. This restriction on the wave structure can be satisfied for generic hydrographs, but various possible situations can cause this assumption to break down, such as adopting multiple peaked hydrographs or stepped hydrographs alternating fast and slow rises. Another example is applying the solution to compound channels, in which the relation of wave speed to discharge is not monotonously increasing (Jacovkis and Tabak, 1996; Chen, 2017). In these circumstances or when there is a non-zero initial discharge in the channel, more complicated wave structures may result (Hunt, 1995; Jacovkis and Tabak, 1996; Stilmant et al., 2018). Solutions with no shock, one internal shock, or multiple shocks may develop, instead of the above continuous wave with a single shock at the wave front.

If its conditions of validity are satisfied, the solution can be used to derive other information of great interest. In particular, it can be used to characterize the flood attenuation behavior. In the kinematic wave description, attenuation of the maximum discharge starts only when the characteristic carrying the peak discharge $Q_P$ reaches the shock at the wave front (Capart, 2013). At the upstream boundary, let the single peaked hydrograph reach its peak discharge $Q_P = \max(Q_B(t))$ at time to peak $T_P = \arg\max(Q_B(t))$. The maximum discharge $Q_{\max}(x)$ that will be experienced at each position $x$ down the valley is then given by

$$Q_{\max}(x) = \begin{cases} Q_P, & x \leq x_T \\ Q_F(x), & x \geq x_T \end{cases}, \qquad (26)$$

where $Q_F(x)$ is the front discharge at position $x = x_F(t_0)$, given by

$$Q_F(x) = Q_B(t_0), \qquad (27)$$

and where $x_T$ denotes the transition position where the maximum discharge starts to decrease, given by

$$x_T = x_F(T_P) = \frac{\alpha m^{1/\alpha}}{\alpha - 1} \frac{V_B(T_P)}{Q_P^{1/\alpha}} = \frac{\alpha}{\alpha - 1} \frac{V_B(T_P)}{A_P}. \qquad (28)$$

This result of great practical importance describes the variation down valley of the maximum discharge experienced during the flood, which in turns controls the extent of flooding. This depends on the valley properties (exponent $\alpha$ and coefficient m), but also on the hydrograph properties including of course the value of the peak discharge $Q_P$, but also the time to peak $T_P$ and the shape of the hydrograph. For a hydrograph that peaks fast, the maximum discharge will start to attenuate at short distances down the valley. For a slowly evolving hydrograph, by contrast, the peak discharge will be experienced over long distances down the valley.

## 3.2. Illustration and validation

To illustrate and validate the above results, we apply them to two special cases for which closed form analytical solutions are available: the sudden dam-break solution derived by Hunt (1984), and the gradual dam breach solution obtained by Capart (2013), both pertaining to U-shaped valleys ($\alpha = 3/2$). The corresponding upstream hydrographs are illustrated in Fig. 4.

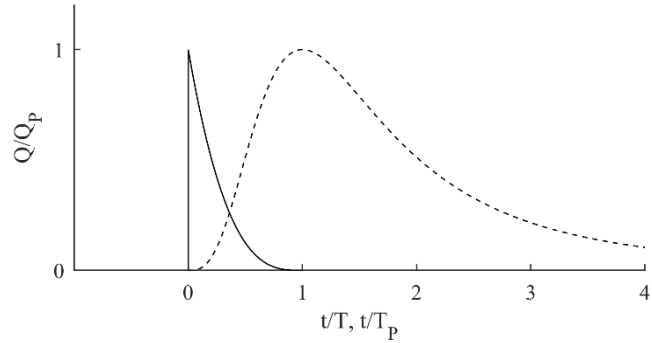

**Fig. 4.** Analytical dam breach hydrographs. Solid line: sudden dam-break; dash line: gradual dam breach.

In Fig. 5, we illustrate the two solutions by plotting characteristic and shock paths in the $(x,t)$ plane, and the corresponding flood attenuation curves down valley. For the sudden dam-break case, the upstream discharge hydrograph is given by

$$Q_B(t_0) = \begin{cases} Q_P(1-t_0/T)^3, & 0 \leq t_0 \leq T \\ 0, & t_0 > T \end{cases}, \quad (29)$$

where the time to peak in this case is $T_P = 0$ (the dam is assumed to vanish instantaneously), and $T$ is the time taken for the lake to drain completely. Substituting this boundary condition into Eq. (24) and (25) yields

$$t_F = \frac{T}{2}\left(\frac{1}{(1-t_0/T)^3} - 3\left(1-\frac{t_0}{T}\right) + 2\right), \quad x_F = \frac{Q_P T}{4 A_P}\left(\frac{1}{(1-t_0/T)^2} - (1-t_0/T)^2\right), \quad (30)$$

which coincide with the results of Hunt (1984) obtained by a different method. Since the upstream discharge attains its peak value instantaneously ($T_P = 0$), the maximum discharge also immediately attenuates with distance down the valley ($x_T = 0$).

For the gradual dam breach case, the upstream hydrograph is given by

$$Q_B(t) = Q_P \frac{8(t/T_P)^3}{(1+(t/T_P)^2)^3},\tag{31}$$

where $T_P$ is the time to peak, and the discharge decreases to zero asymptotically. Substituting Eq. (31) into Eq. (24) and (25) yields for the wave front path the results

$$t_F = \frac{t_0}{2}\left(3+\left(\frac{t_0}{T_P}\right)^2\right), \quad x_F = \frac{3Q_P T_P}{2A_P}\left(\frac{t_0}{T_P}\right)^2.\tag{32}$$

Likewise, Eq. (28) yields the following result for the transition position where discharge attenuation starts:

$$x_T = \frac{3Q_P T_P}{2A_P}.\tag{33}$$

These results are identical with those of Capart (2013). For the two special cases, therefore, our more general solutions coincide with earlier results. The plots of Fig. 5 further highlight how suddenly and gradually rising hydrographs produce very different behaviors. For the sudden dam-break case (left panels), the wave front path decelerates immediately as it travels down valley, and the maximum discharge likewise attenuates immediately. For the gradual dam breach case, by contrast (right panels), the wave front path first accelerates, then decelerates down valley. It takes some time for to hydrograph to reach its peak value, then some more time before the characteristic carrying the peak discharge reaches the wave front shock. As a result, the peak discharge is experienced over some distance before the maximum discharge starts to attenuate down valley. As we examine in the next section the more complex behaviors produced in non-uniform valleys, these results for uniform valleys will provide a valuable baseline.

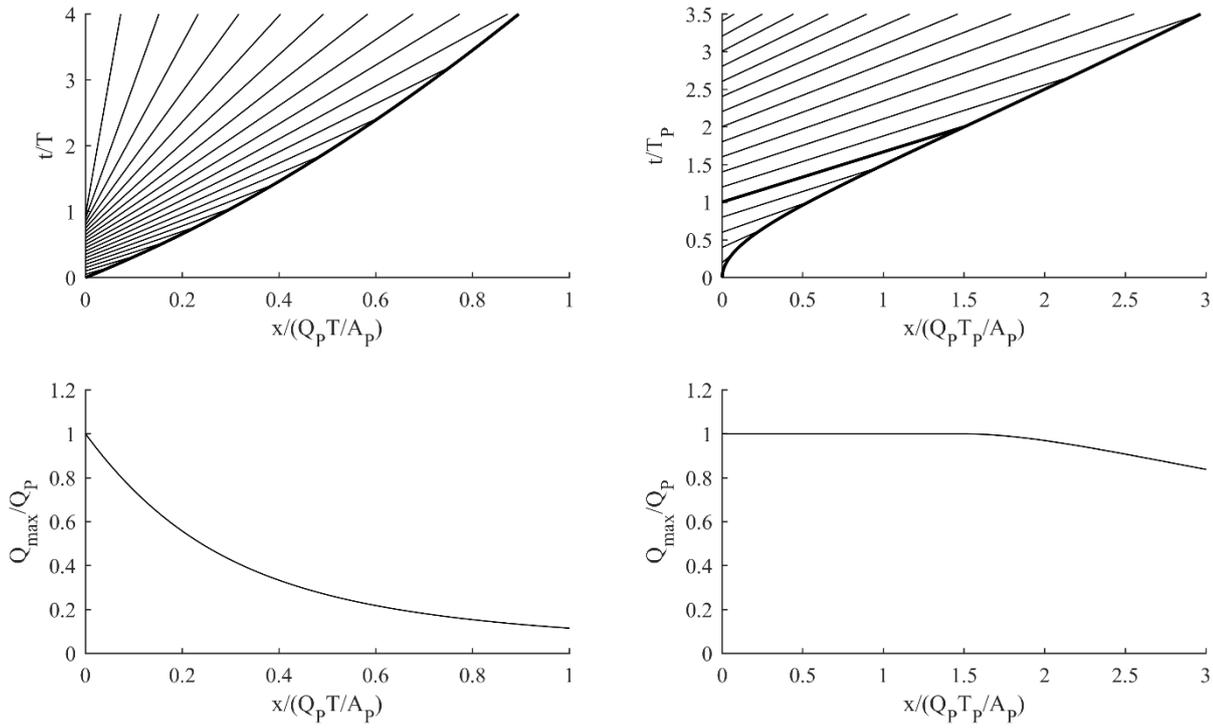

**Fig. 5.** Solutions for the analytical dam breach cases. Top: characteristics and shock paths in the $(x,t)$ plane; bottom: corresponding flood attenuation curves down valley.

## 4. Solutions for non-uniform valleys

### 4.1. Solution Method

For non-uniform valley, we need to solve Eq. (1) with $A$ dependent on both $Q$ and $x$. This is difficult in general, but simplified considerably if the relation $A(Q,x)$ is separable, i.e. if we can write

$$A(Q,x) = f_1(Q) f_2(x). \qquad (34)$$

For rating curves in the power law form of Eq. (10), this is the case with

$$f_1(Q) = Q^{1/\alpha}, \quad f_2(x) = m(x)^{-1/\alpha}. \qquad (35)$$

where the exponent $\alpha = 3/2$ or $5/4$ for U and V-shaped valleys, respectively, and $m(x)$ is the coefficient describing the variation with distance of channel properties like width, slope and friction factor. When the relation $A(Q,x)$ is separable, it becomes possible to homogenize channel variations by making a change of variable and rescaling the distance $x$. Applied to the time $t$ instead of the distance $x$, a similar approach was earlier used to solve the diffusion equation with time-varying diffusion coefficient (Crank, 1975; Capart, 2013). For kinematic waves in non-uniform valleys, we obtain the rescaled distance variable $\xi(x)$ in the following way. Assuming $A(Q,x)$ is separable, the continuity equation, Eq. (1), can be written

$$f_1(Q)\frac{\partial f_2(x)}{\partial t} + f_2(x)\frac{\partial f_1(Q)}{\partial t} + \frac{\partial Q}{\partial x} = f_2(x)\frac{\partial f_1(Q)}{\partial t} + \frac{\partial Q}{\partial x} = 0. \quad (36)$$

To simplify the equation, we then define

$$\frac{d\xi}{dx} = \frac{f_2(x)}{f_2(x_0)}, \quad (37)$$

where $x_0$ is an arbitrary reference position used to normalize channel properties. Eq. (36) can then be rewritten

$$\frac{\partial}{\partial t}(f_1(Q)f_2(x_0)) + \frac{\partial Q}{\partial \xi} = 0, \quad (38)$$

where

$$f_1(Q)f_2(x_0) = A(Q,x_0) = A_0(Q). \quad (39)$$

For non-uniform valleys, the continuity equation can then be written in the form

$$\frac{\partial A_0(Q)}{\partial t} + \frac{\partial Q(\xi,t)}{\partial \xi} = 0. \quad (40)$$

For uniform valleys, the same equation can be written

$$\frac{\partial A_0(Q)}{\partial t} + \frac{\partial Q(x,t)}{\partial x} = 0. \quad (41)$$

Except for the change of variable from $x$ to $\xi$, Eq. (40) is identical to Eq. (41). The equation for non-uniform valleys therefore reduces to the equation for uniform valleys when the problem is mapped from the $(x,t)$ plane to the $(\xi,t)$ plane. Solutions for non-uniform valleys can therefore be obtained from solutions for uniform valleys by simply substituting $\xi(x)$ for $x$ in the corresponding formulas. This rescaled distance $\xi(x)$ can be obtained by integration from Eq. (37), yielding

$$\xi(x) = \int_0^x \frac{f_2(x')}{f_2(x_0)} dx', \tag{42}$$

where $x'$ is a variable of integration. To obtain the wave speed $\lambda$, we can calculate

$$\frac{\partial A}{\partial Q} = \frac{\partial f_1(Q) f_2(x_0)}{\partial Q} = f_2(x_0) \frac{\partial f_1(Q)}{\partial Q}, \tag{43}$$

and therefore

$$\lambda(x,t) = \frac{1}{\partial A / \partial Q} = \frac{1}{f_2(x_0) \dfrac{\partial f_1(Q)}{\partial Q}} = g_1(Q) g_2(x), \tag{44}$$

hence the relation for the wave speed $\lambda$ is also separable. As a consequence,

$$\frac{\lambda_1(Q, x_1)}{\lambda_2(Q, x_2)} = \frac{g_1(Q) g_2(x_1)}{g_1(Q) g_2(x_2)} = \frac{g_2(x_1)}{g_2(x_2)}. \tag{45}$$

In non-uniform valleys, the speed of a flood wave will accelerate or decelerate by a factor dependent on its position $x$, but not on its discharge $Q$. Like light propagating in a non-homogenous medium, flood waves in non-uniform valleys alter their speed $\lambda$ with position, hence they experience refraction due to variations in channel properties. By using the rescaled distance $\xi(x)$, however, we can make these wave speed variations disappear.

Restricting our attention to rating curves in power law form, the relation for the flow area becomes

$$A(Q, x) = f_1(Q) f_2(x) = Q^{1/\alpha} m(x)^{-1/\alpha}. \tag{46}$$

The change of variable needed to homogenize the channel properties is then, in differential form,

$$\frac{d\xi}{dx} = \frac{m_0^{1/\alpha}}{m(x)^{1/\alpha}}, \tag{47}$$

and in integral form

$$\xi(x) = \int_0^x \frac{m_0^{1/\alpha}}{m(x')^{1/\alpha}} dx'. \tag{48}$$

Note from Eq. (46) that the exponent $\alpha$ must be constant for the relation to be separable and for the rescaling to work. The width, slope and friction factor can change freely with distance, but the type of cross-section (U or V-shaped) must remain the same. If the valley contains segments of different shapes, the approach can still be used, but must be applied piecewise to segments of constant exponent $\alpha$. For each segment, the solution must be calculated using the outflow hydrograph from the previous segment as upstream boundary condition.

Segment by segment, or for the whole valley if the channel type does not change, the above rescaling makes it simple to route dam-break floods down non-uniform valleys. The flood wave behavior, wave front propagation, and maximum discharge attenuation can therefore be calculated as if the valley was uniform. The results can then be mapped back to the true distance $x$ using the reciprocal relation $x(\xi)$. The results also clarify the effects of channel properties on flood attenuation. Where the river slope $S$ is small, the friction factor $f$ large, the width $b$ large (for U-shaped valley) or the bank slope $m_v$ mild (for V-shaped valleys), the rating curve coefficient $m$ will be small and the rate $d\xi/dx$ large, locally causing a more rapid attenuation of the maximum flood discharge. In accordance with intuition, therefore, flood attenuation will be more rapid in wide valleys of mild inclination than in narrow, steep gorges. These effects will be illustrated below for both idealized and real cases.

## 4.2. Illustration and validation

To illustrate and validate the above results, we consider three hypothetical non-uniform valleys A, B, and C. Valley A and B are U-shaped valleys ($\alpha_A = \alpha_B = 3/2$) of length $L$ with only three known cross sections located at $0.5L$, $1.5L$ and $2.5L$. There the channel widths are respectively 4/5, 1/5 and 2 times the average width $b_0$ of the three cross-sections. The slope $S$ and friction factor $f$ are assumed constant. The two valleys differ in how the width is assumed to vary between the known cross sections. Segments of piecewise constant width are assumed for valley A, whereas linear variations are assumed for valley B. For valley A, the exponent $\alpha$ and the valley width function $b(x)$ are therefore given by

$$\alpha_A = \frac{3}{2}, \quad \frac{b_A(x)}{b_0} = \begin{cases} 4/5, & 0 \leq x < L \\ 1/5, & L \leq x < 2L \\ 2, & 2L \leq x < 3L. \end{cases} \tag{49}$$

For valley B, they are

$$\alpha_A = \frac{3}{2}, \quad \frac{b_A(x)}{b_0} = \begin{cases} 3/5(1.5 - x/L) + 1/5, & 0 \leq x < 1.5L \\ 9/5(x/L - 1.5) + 1/5, & 1.5L \leq x < 3L. \end{cases} \tag{50}$$

The two assumed width variation functions are illustrated in Fig. 6.

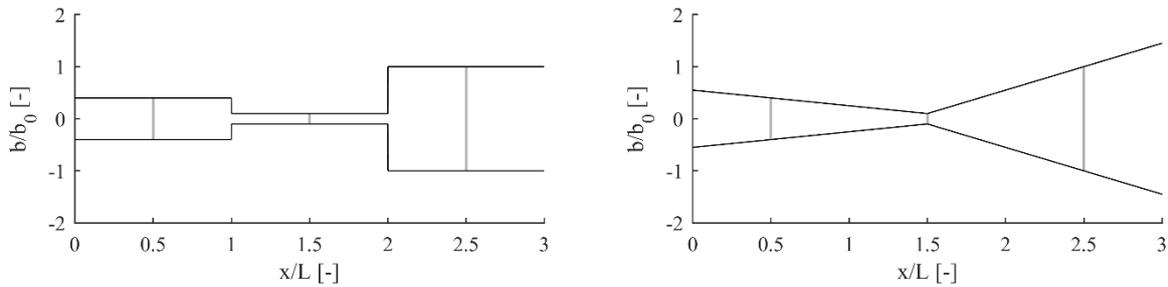

**Fig. 6.** Assumed width functions for Valley A (piecewise constant) and valley B (piecewise linear). Gray lines indicate the positions and widths of the three known cross-sections.

Valley C is a mixed valley composed of two segments. The valley is V-shaped over the upstream segment, and U-shaped over the downstream segment, both having length $L$. The corresponding exponents $\alpha$ and coefficients $m(x)$ are given by

$$\alpha_C = \begin{cases} 5/4 \\ 3/2 \end{cases}, \quad \frac{m_C(x)}{m_0} = \begin{cases} 1 - x/2L, & 0 \leq x < L \\ 1/10(1.5 - x/2L), & L \leq x < 2L. \end{cases} \tag{51}$$

The properties of the three valleys are listed in Table 1.

**Table 1.** Properties of the three valleys.

| Valley | $\alpha$ | $m(x)/m_0$ | x |
|---|---|---|---|
| A | 3/2 | $(4/5)^{-2}$ | $0 \leq x < L$ |
|   |     | $(1/5)^{-2}$ | $L \leq x < 2L$ |
|   |     | $(2)^{-2}$ | $2L \leq x \leq 3L$ |
| B | 3/2 | $(3(1.5 - x/L)/5 + 1/5)^{-2}$ | $0 \leq x < 1.5L$ |
|   |     | $(9(x/L - 1.5)/5 + 1/5)^{-2}$ | $1.5L \leq x \leq 3L$ |
| C | 5/4 | $1 - x/2L$ | $0 \leq x < L$ |
|   | 3/2 | $(1.5 - x/2L)/10$ | $0 \leq x \leq 2L$ |

Initially, all three valleys are assumed to be dry. At their upstream ends, we consider two different discharge hydrographs. The first, assumed to result from a sudden dam failure, is given by

$$Q_B(t) = Q_P(1 + t/T)^{-3}, \tag{52}$$

where $Q_P$ is the peak discharge, attained immediately after failure, and $T$ is a time scale for the falling limb of the hydrograph. The second is the gradual dam breach hydrograph of Eq. (31), considered in the previous section. To check the analytical results, they will be compared to numerical solutions computed using the finite volume scheme described in Appendix A.

For Valley A and B, we can use Eq. (24) and (25) to obtain flood solutions in the $(\xi,t)$ plane, as shown in Fig. 7a,c and Fig. 8a,c. In the $(\xi,t)$ plane, the results are identical to solutions for uniform valleys of constant width $b_0$ for the same hydrographs. These can then be mapped to the $(x,t)$ plane to obtain solutions in the true coordinates, as shown in Fig. 7b,d and Fig. 8b,d. In the $(\xi,t)$ plane, all characteristic paths are straight rays, which do not deviate until they impinge upon the curved path of the wave front. In the $(x,t)$ plane, by contrast, the characteristic paths are refracted by variations in channel properties. For valley A, these variations are abrupt, causing the rays and wave front to bend sharply at transitions. For valley B, by contrast, variations are gradual, causing the rays to curve and the wave front to bend continuously. In both cases, the characteristic and wave front speeds accelerate in the middle reaches, where the channel narrows, illustrating the behavior expected when flood waves travel through a narrow gorge.

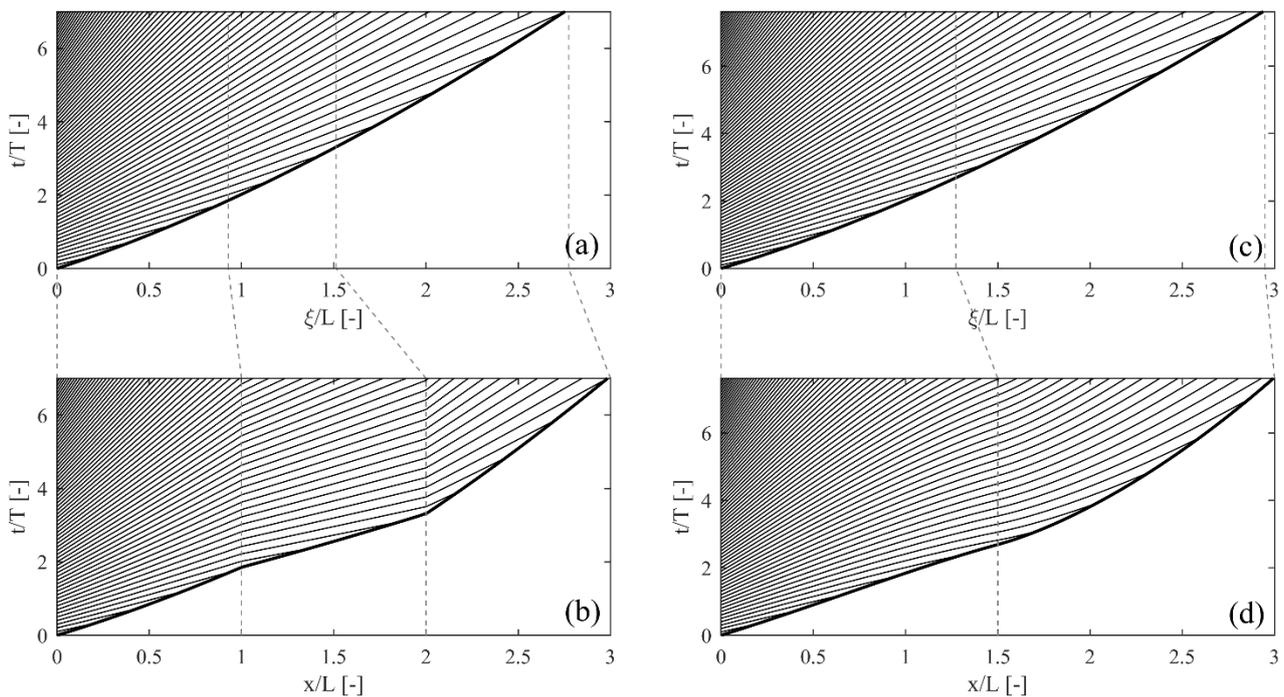

**Fig. 7.** Kinematic wave solution for the routing of a sudden dam-break wave through valleys A and B: (a) results for valley A in the $(\xi,t)$ plane, where the valley is experienced as uniform; (b) results for valley A in the $(x,t)$ plane, where refraction due to variations in channel width is apparent; (c) results for valley B in the $(\xi,t)$ plane; (d) results for valley B in the $(x,t)$ plane. Thin lines represent characteristic rays (which are also level sets of the discharge), the bold line is the path of the wave front, and thin dashes denote valley transitions.

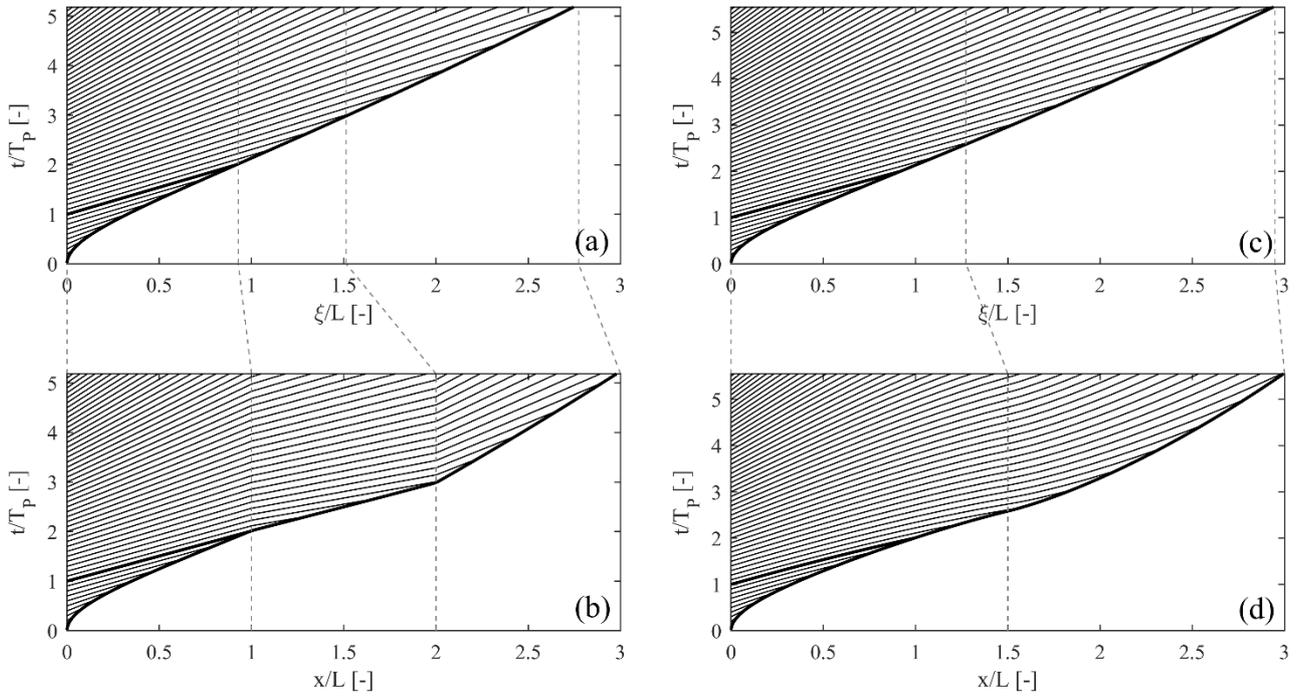

**Fig. 8.** Kinematic wave solution for the routing of a gradual dam-break wave through valleys A and B: (a) results for valley A in the $(\xi, t)$ plane, where the valley is experienced as uniform; (b) results for valley A in the $(x, t)$ plane, where refraction due to variations in channel width is apparent; (c) results for valley B in the $(\xi, t)$ plane; (d) results for valley B in the $(x, t)$ plane. Thin lines represent characteristic rays (which are also level sets of the discharge), bold line are the paths of the peak discharge and wave front, and thin dashes denote valley transitions.

In Fig. 9, we compare analytical and numerical results for valleys A and B. For clarity, the corresponding valley width variations are shown in Fig. 9a, d. Results for the wave front path are plotted in Fig. 9b, e. The analytical (thin lines) and numerical results (dots) are found to agree closely with each other, validating our approach. Finally, results for the flood attenuation behavior are plotted in Fig. 9c, f. As observed earlier in uniform valleys, for sudden dam-break waves the maximum discharge attenuates immediately. For gradual dam-break waves, by contrast, the maximum discharge remains equal to the peak value for some distance down the valley, before gradually decreasing. The pace at which this process unfolds, however, is influenced by variations in valley geometry. Attenuation is delayed where the valley is narrow, and accelerated where the valley is wide. For gradual dam-break waves, attenuation starts at a given value of the rescaled variable $\xi_T$, but the corresponding true transition position $x_T$ can be advanced or delayed depending on the channel properties experienced along the way by the flood. Again, the analytical (lines) and numerical results (dots) are found to agree closely with each other. For non-uniform valleys of a single type, our proposed analytical approach is therefore demonstrated to provide reliable results.

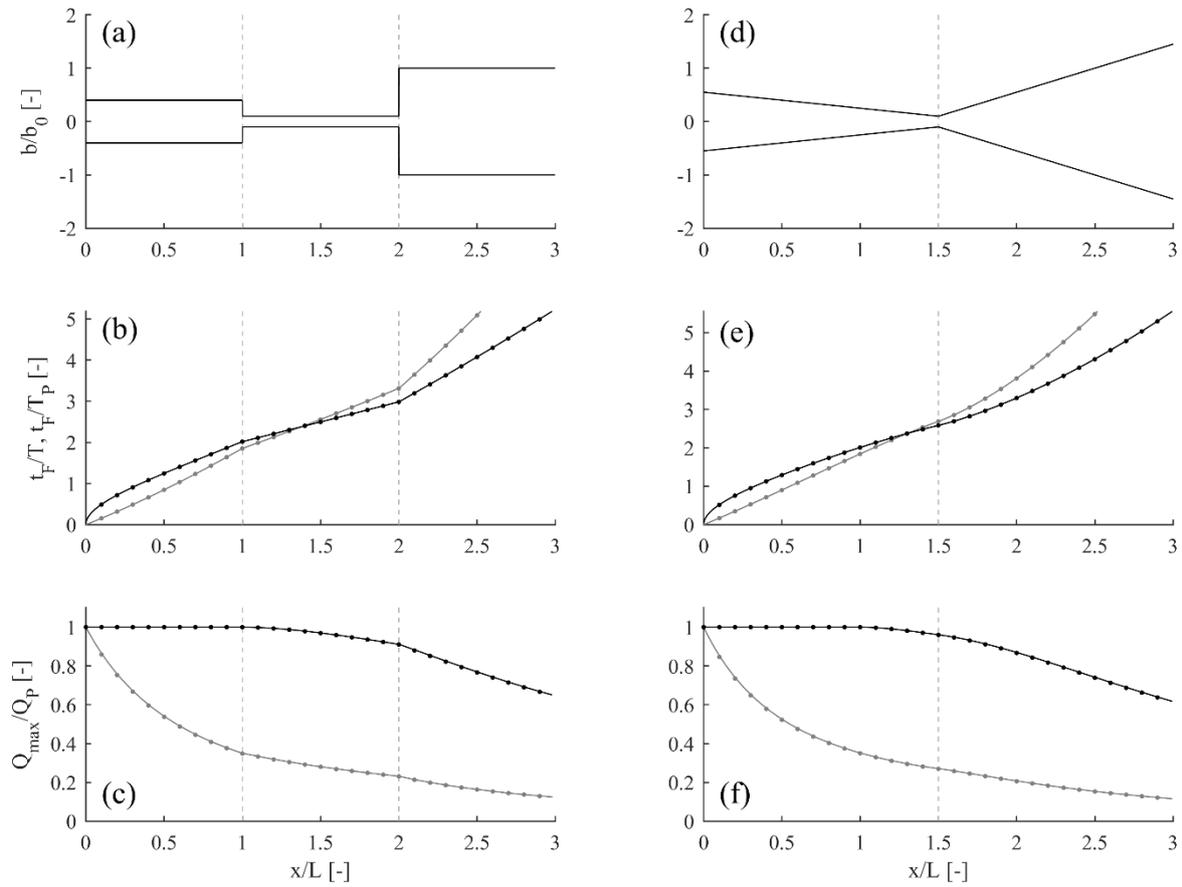

**Fig. 9.** Comparison of analytical (lines) and numerical solutions (dots) for Valley A and B, sudden and gradual hydrographs: (a) width variation of Valley A; (b) wave front path (arrival time) in Valley A, for sudden (gray) and gradual dam-breaks (black); (c) attenuation of the maximum discharge in Valley A, for sudden (gray) and gradual dam-breaks (black); (d) width variation of Valley B; (e) wave front path (arrival time) in Valley B; (f) attenuation of the maximum discharge in Valley B.

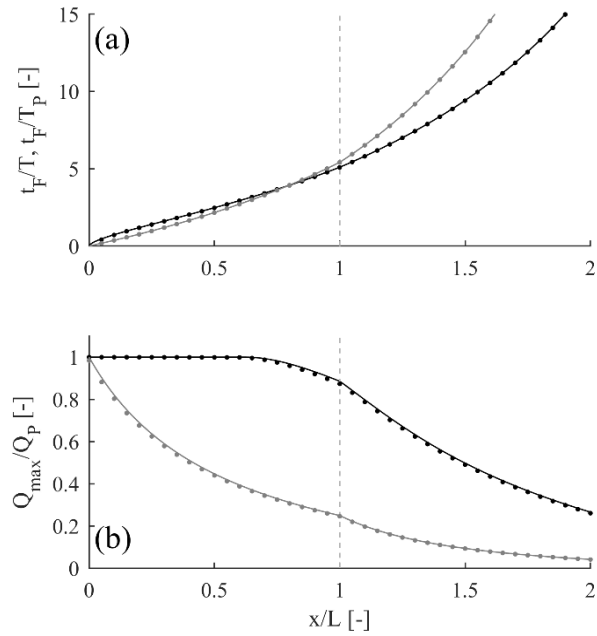

**Fig. 10.** Comparison of numerical solutions (dots) and analytical solutions (lines) for Valley C: (a) wave front path (arrival time) for sudden (gray) and gradual dam-break (black); (b) attenuation of the maximum discharge for sudden (gray) and gradual dam-break (black).

To check that the approach can also be used for mixed valleys, Fig. 10 compares analytical and numerical results for Valley C. In this case, solutions must be computed in two steps. Flood waves are first routed through the upstream V-shaped valley segment using the same method as before. The outflow hydrograph at its downstream end is then used as boundary condition for the downstream U-shaped valley segment, through which the flood wave is routed by the same method. The analytical (lines) and numerical results (dots) are again in close agreement, validating our method for the case of mixed valleys.

Note here that we are free to assume abrupt variations for hypothetical cases A and C only in the context of the kinematic wave approximation. In well-resolved dynamic simulations, such abrupt variations would induce oscillations due to wave reflections (Begnudelli and Sanders, 2007), not described by the kinematic model.

## 5.  Application to field historical cases

To test the proposed approach for real cases, we simulate in this section three well-documented field events: the 2008 breaching failure of the Tangjiashan landslide dam, the 1976 piping failure of Teton Dam, and the 1959 sudden failure of Malpasset Dam. Due to their different modes of failure, their hydrographs feature slow, fast, and abrupt rises to their peak values. They also differ in valley geometry, level of available topographic detail, and type of measured data available for comparison. In Appendix B, we describe the method we use to reconstruct the discharge hydrographs released by the failed dams, based on storage curves, assumptions about the failure process, and contemporary records. Other assumptions and the simulation results are described separately for each case in the following three sections.

### 5.1. Breaching failure of the Tangjiashan landslide dam

The Tangjiashan natural dam was formed on 12 May 2008 when a landslide triggered by the Wenchuan Earthquake caused the damming of Tongkou River (see Fig. 1). The dam breached due to overtopping on 10 June 2008, approximately one month later. The case is an excellent example of successful engineering intervention: the time available between the damming event and the dam breach was exploited to excavate the dam crest to a reduced height, evacuate downstream areas, and instrument the site. As a result, the dam breach flood was much less severe then it could have been, and exceptionally well-documented (Liu et al. 2009, 2010).

To calculate the upstream discharge hydrograph, we use the dam failure model described in Appendix B. The model uses the measured lake storage curve upstream of the dam (Fan et al., 2012, Fig. 11a), and is calibrated by matching the observed peak discharge $Q_P$ and time to peak $T_P$. In Fig. 11b, we compare our simulated hydrograph with those deduced by Liu et al. (2010) from measured data. The two are in reasonable agreement, but our simulated hydrograph declines somewhat slower than the observations. Whereas the RMSE between the two measured hydrographs is 1141 cms, the root-mean-square errors (RMSE) between the simulated hydrograph and the two measured hydrographs are computed to be 709 cms and 1042 cms, respectively. The discrepancy between the simulated and measured hydrographs is therefore no larger than the difference between the measurements.

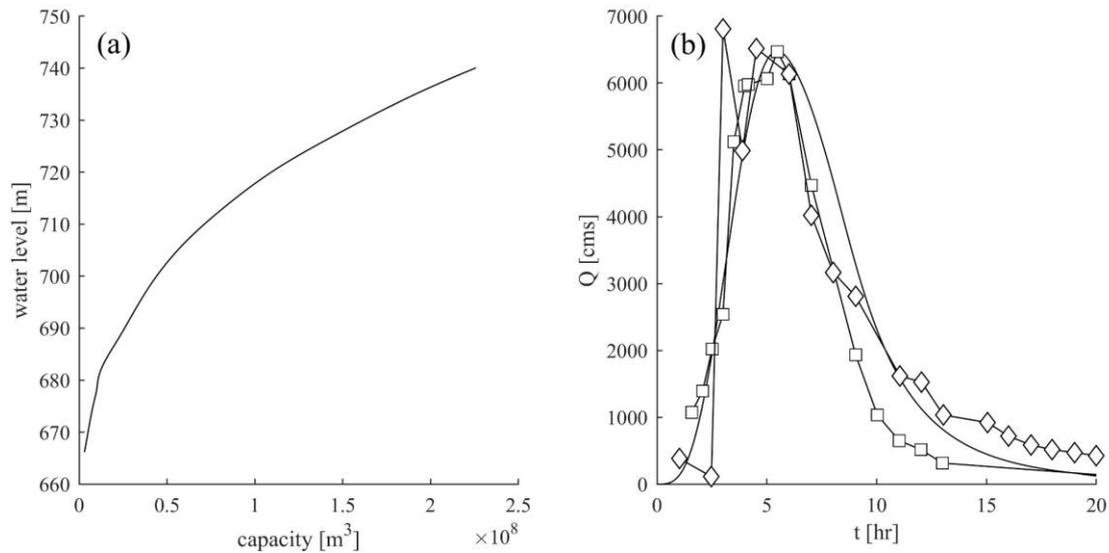

**Fig. 11.** Assumed upstream discharge hydrograph for the Tangjiashan dam breach flood: (a) storage curve of the Tangjiashan barrier lake (Fan et al., 2012); (b) comparison of our simulated hydrograph (solid line) with the measured hydrographs deduced from measured velocities and depth (squares) and from lake level drawdown (diamonds) (Liu et al., 2010).

Downstream of the dam, cross section data are available at the Beichuan, Tongkou, and Mianyang gauging stations, located respectively 5.3, 27.3, and 74.9 km downstream of the dam, and the valley is approximately U-shaped (see Fig. 1). The corresponding channel widths are listed in Table 2, and vary between 100 m and 300 m. Note that the calibrated friction factors represent an effective resistance, not limited to just friction along the valley perimeter. This effective resistance may account for other energy dissipation terms, such as oblique hydraulic jumps (Begnudelli and Sanders, 2007). Fitting errors between the power-law rating curves and the stage-discharge data are given in Fig. 12 and Table 3. Since the discharge measurements used to produce the rating curves are themselves uncertain, the calculated fitting error provides only a partial estimate of the rating curve uncertainty. Furthermore, because cross sections are known only at three locations, some assumption is needed to define a complete valley geometry. As illustrated in Fig. 13, we will compare two different assumptions for the width profile $b(x)$: piecewise constant and linearly varying. Channel slopes are assumed equal to 0.0034 and 0.0016, respectively, in the upstream valley and downstream plain. We will also compare our results with those of Capart (2013), who assumed a uniform valley of constant effective width and slope (see Table 2).

**Table 2.** Flood routing parameters for the three cross-section of Tongkou River.

|  | Parameter (units) | Value | Parameter (units) | Value |
|---|---|---|---|---|
| Beichuan station | $x_A$ (km) | 5.3 | $z_A$ (m) | 611.9 |
|  | $b_A$ (m) | 165 | $f_A$ (-) | 0.103 |
| Tongkou station | $x_B$ (km) | 27.3 | $z_B$ (m) | 536.3 |
|  | $b_B$ (m) | 100 | $f_B$ (-) | 0.180 |
| Mianyang station | $x_C$ (km) | 74.9 | $z_C$ (m) | 459.5 |
|  | $b_C$ (m) | 290 | $f_C$ (-) | 0.057 |
| Effective uniform river valley | $S_V$ (m/m) | 0.0024 | $f_V$ (-) | 0.086 |
|  | $b_V$ (m) | 185 |  |  |

**Table 3.** Errors between simulated results and measurements in field.

|  |  | Dam site | Beichuan station | Tongkou station | Mianyang station |
|---|---|---|---|---|---|
| Simulated discharge | $Q_P$ error [%] | +0.7[a]/-4.0[b] | -0.2 | +6.7 | +6.8 |
|  | RMSE [cms] | 709[a]/1042[b] | 981 | 894 | 1347 |
|  | RMSE/$Q_P$ [-] | 10.8[a]/16.0[b] | 15.1 | 13.7 | 20.6 |
| Measured discharge | RMSE [cms] | 1141 | - | - | - |
|  | RMSE/$Q_P$ [%] | 17.5 | - | - | - |
| Simulated water depth | $H_{max}$ error [%] | - | +6.6 | -7.0 | +6.4 |
|  | RMSE [m] | - | 1.64 | 2.21 | 1.6 |
|  | RMSE/$H_{max}$ [%] |  | 18.4 | 17.7 | 26.1 |
| Measured water depth (uncertainty due to rating curve) | RMSE [m] | - | 0.07 | 0.3 | 0.17 |
|  | RMSE/$H_{max}$ [%] | - | 0.7 | 0.2 | 0.3 |

[a] Errors compared to the dam breach hydrograph estimated from observed velocity and depth.

[b] Errors compared to the dam breach hydrograph estimated from observed lake level.

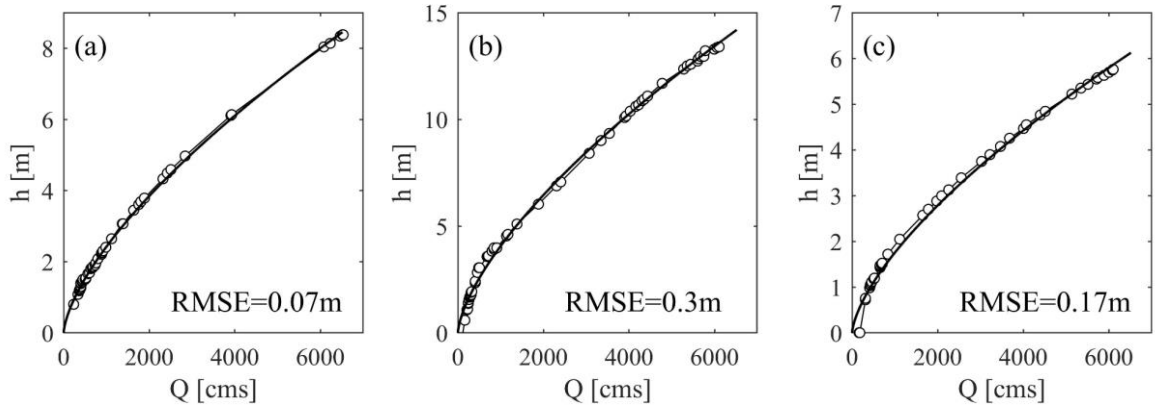

**Fig. 12.** Rating curves of the gauge stations (solid lines: fitted rating curve; circles: measured data). (a) Beichuan station; (b) Tongkou station; (c) Mianyang station.

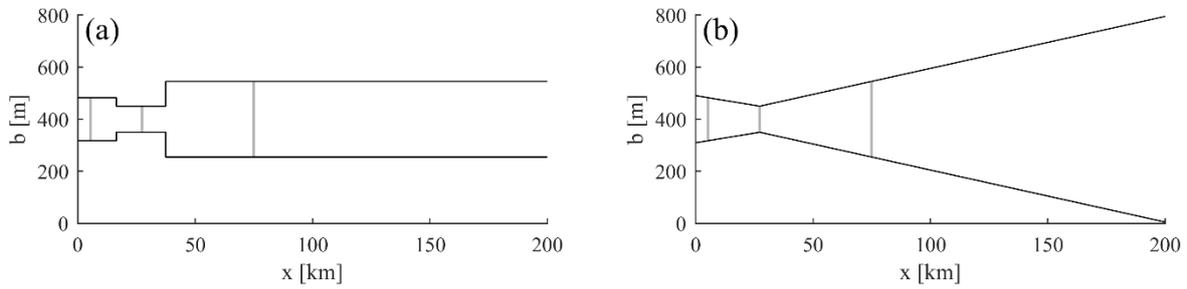

**Fig. 13.** Assumptions for the valley width: (a) piecewise constant width profile; (b) linearly varying profile.

Based on these input data, the analytical method described in the previous section can be used to route the dam breach flood. Results are first calculated in the $(\xi,t)$ plane, then converted to the $(x,t)$ plane using the variable rating curve coefficient $m(x)$ estimated from the assumed friction factor, width profile and slope profile. The simulated results obtained in this way are shown in Fig. 14, and compared with the field measurements. The errors between the field measurements and the simulated results are presented in Table 3.

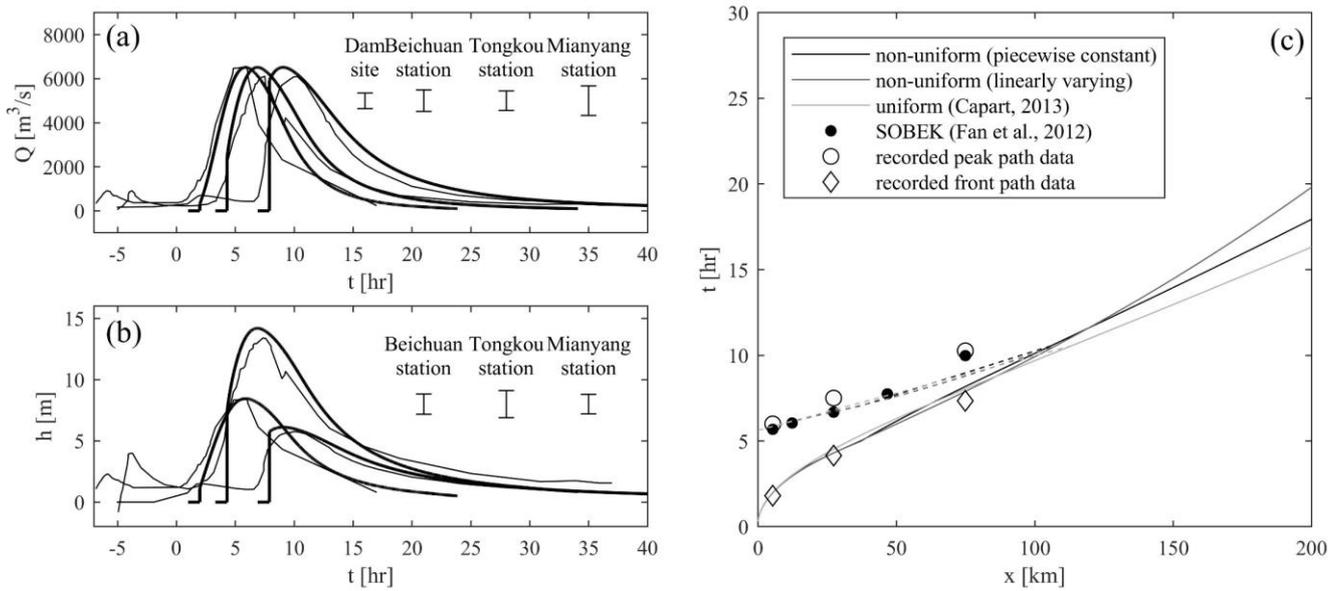

**Fig. 14.** Comparison of simulated and measured results for the Tangjiashan case: (a) measured (thin lines) and simulated (thick lines) discharge hydrographs at the three gauging stations. Error bars represent the root-mean-square error (RMSE) between simulated and measured hydrographs; (b) measured (thin lines) and simulated (thick lines) depth hydrographs at the three stations. Error bars represent the RMSE between simulated and measured depth hydrographs; (c) measured (hollow symbols) and simulated paths (lines) of the wave front (solid lines) and discharge peak (dashed lines). The black dots are the peak discharge arrival times simulated by Fan et al. (2012) using Sobek.

In Fig. 14a, we first plot the discharge hydrographs at the three gauging stations. At all three stations, the maximum discharge predicted by the kinematic model has not yet started to attenuate. Because the hydrograph rises slowly to its peak value, an even greater distance is needed before attenuation occurs. The rising limb of the simulated discharge hydrograph, nevertheless, has become steeper by the time it reaches Mianyang station. Likewise, the field measurements show minimal attenuation of the peak, and some degree of steepening of the rising limb. The discontinuous shock predicted by the kinematic model at the wave front, however, is less abrupt in the measured data. This causes the increased errors at the three stations (981, 894, and 1347 cms RMSE, respectively) and is likely due to momentum effects neglected in the kinematic approximation (Singh, 2002; Bohorquez, 2010; Warnock et al., 2013). Besides, the discrepancy for the dam breach hydrograph (709-1042 cms RMSE) will also contribute to errors in the simulated discharges.

Corresponding results for the depth hydrographs are shown in Fig. 14b. Due to variations in width, very different responses are observed at the three gauging stations. At the upstream and downstream stations, where the valley is wide, the water depths induced by the flood do not exceed 8 m. At the middle station, by contrast, where the valley is narrow, the water depth goes up to more than 12 m for approximately the same peak discharge. For these effects due to the non-uniform valley geometry, the simulated results agree reasonably well with the measured data. The errors on water depth can be partly attributed to the uncertainty of the rating curves, as mentioned above. They may also be due to errors in the discharge hydrographs, or to deviations from the one-to-one relationship between stage and discharge assumed in the kinematic wave approximation. In Fig. 14c, finally, we show results for the paths of the wave front and peak discharge. Simulated results assuming a non-uniform valley differ significantly from those of Capart (2013), calculated assuming a uniform valley of average properties. They agree slightly better with the measured data, and with the two-dimensional simulations conducted by Fan et al. (2012) using the software Sobek. Except at the downstream end, where extrapolation leads to large differences, it matters less whether we use piecewise constant or linearly varying functions to approximate the width variation profile. Overall, reasonably good agreement is obtained between the simulations and measurements. Although they require much less computational effort, our one-dimensional kinematic simulations also compare well with two-dimensional dynamic simulations.

## 5.2. Piping failure of Teton Dam

Built across Teton River, Idaho, this earthen dam failed in 1976 due to piping, causing severe flooding. The U.S. Geological Survey (1976) collected detailed data following this event, including the inundated area and the maximum water levels along a distance of 160 km down valley. As illustrated in Fig. 15, the width of the inundated area varied greatly with distance, making this a suitable case to examine non-uniform valley effects.

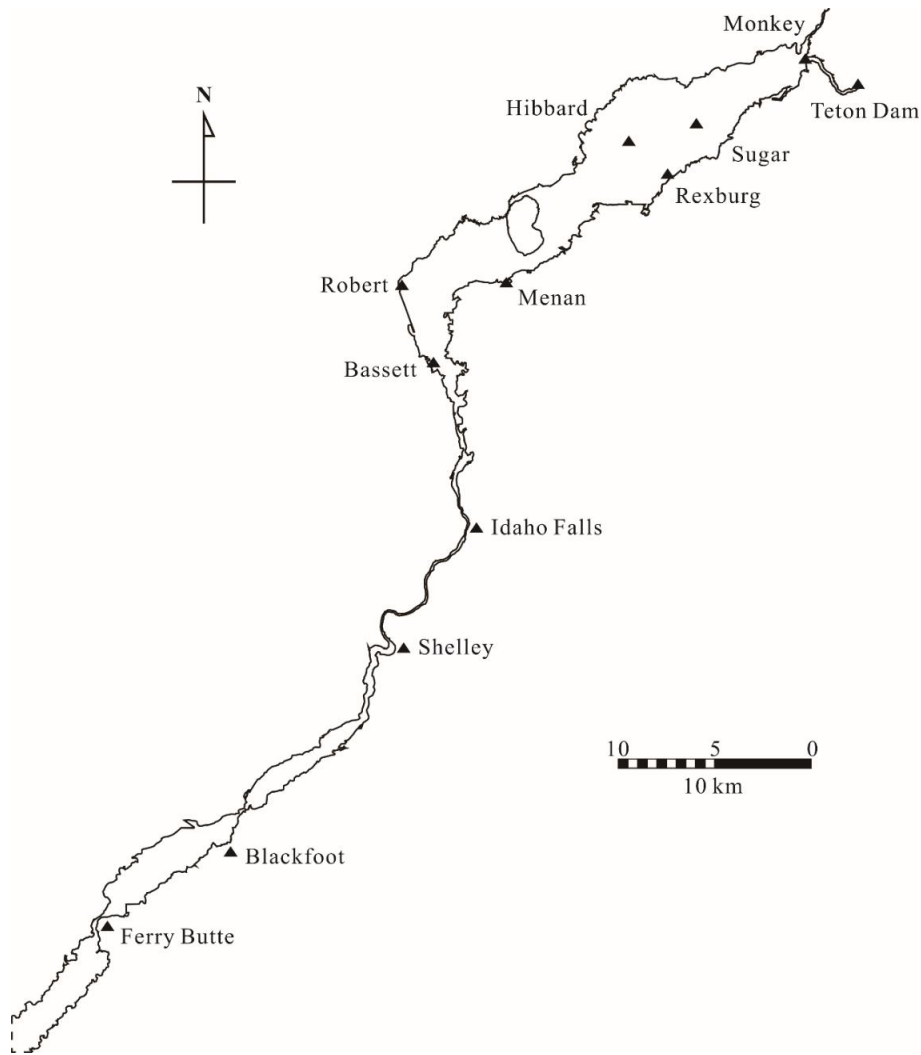

**Fig. 15.** The inundated area due to the Teton dam failure.

To reconstruct the upstream discharge hydrograph, we applied the dam failure model described in Appendix B using the documented reservoir storage curve (Gundlach and Thomas, 1977) shown in Fig. 16a, and model parameters calibrated to match the measured peak discharge $Q_P$ and time to peak $T_P$. As shown in Fig. 16b, the resulting simulated hydrograph agrees reasonably well with the one reconstructed by Balloffet et al. (1982) from field observations. The rising limb of the hydrograph, however, occurs a bit earlier in the simulation.

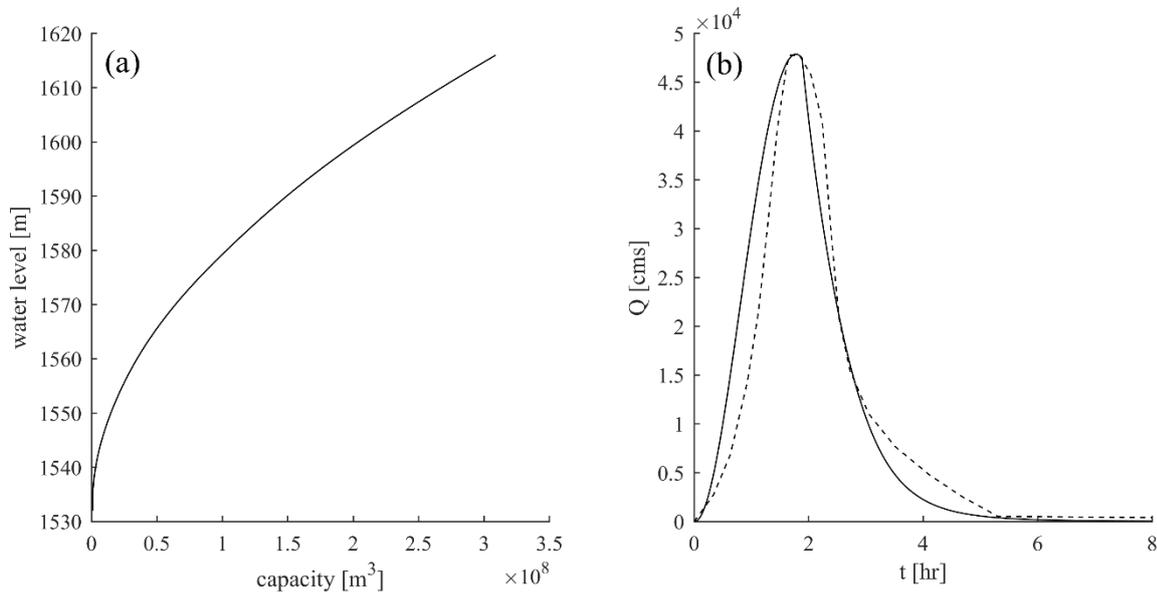

**Fig. 16.** The hydrograph prediction. (a) Storage curve of Teton Reservoir (Gundlach and Thomas, 1977); (b) comparison of predicted hydrograph (solid line) and reconstructed field hydrographs (dash line) for the Teton dam breach (Balloffet et al., 1982).

In the reach between Teton Dam and Monkey and the reach between Bassett and Shelley (0-8 km and 65-110 km in Fig. 17a), the flood was confined in the river channel, which is approximately U-shaped. Hence, rating curves in the two reaches are dominated by longitudinal channel slope $S$ and channel width $b$. By contrast, in other reaches, the flooding greatly exceeded the original cross-section of the river channel and went into the valley with no clear lateral boundaries. We therefore assume a V-shaped valley in these reaches, and hence rating curves that depend both on the longitudinal channel slope $S$ and on the valley side slope $m_v$.

To define the valley topography, we first extract the river long profile (Fig. 17b) and channel slope profile $S(x)$ from the USGS topographic map. The channel width $b$ and the valley side slope $m_v$ respectively for the U and V-shaped reaches are derived from the USGS topographic map using 30 cross sections. The only exception is the valley reach near Rexburg, where the valley features a compound channel. For this reach, the following procedure is used to simplify the cross section into an equivalent V-shaped cross section. First, we calculate the rating curve for the compound channel. Then, we construct a power-law rating curve with exponent $\alpha = 5/4$ by choosing the coefficient m that best fits the flow area $A$ for the two discharge values $Q = 0.2Q_P$ and $0.8Q_P$. The equivalent V-shaped valley side slope $m_v$ is calculated inversely from the fitted rating curve. Therefore, the rating curve coefficient $m(x)$ along the entire valley can be derived by assuming that the channel width of U-shaped valley and the side slope of V-shaped valley vary piecewise linearly. The Darcy friction coefficient $f$ is calculated from the average high water depth, based on the Manning coefficient value $n = 0.06$ s/m$^{1/3}$.

As described in the section 4.2, the analytical solution for mixed valley is applied to the valley segments separately. Flood waves are first routed through the upstream single-shaped valley segment using the same method as before. The outflow hydrograph at its downstream end is then used as upstream boundary condition for the next single-shaped valley segment, through which the flood wave is routed by the same method. The resulting profiles of maximum water level and water depth are shown in Fig. 17b, c, where they are compared with the field records (USGS, 1976) and with simulations conducted using the software HEC-2 (Gundlach and Thomas, 1977). Despite the simplified valley geometry and idealized model assumptions, the water profiles simulated by the kinematic model are in good overall agreement with the surveyed and HEC-2 results. Agreement is poorest in the reach between Idaho Falls and Shelley (about 80-100 km), where the use of the same friction coefficient is less reasonable. The RMSE of the maximum water depth for the entire valley is computed to be 2.63 m, which is close to the average variance of the maximum water depth between the left and right banks (2.17 m) and even smaller than the RMSE of the maximum water depth obtained with the software HEC-2 (3.27 m).

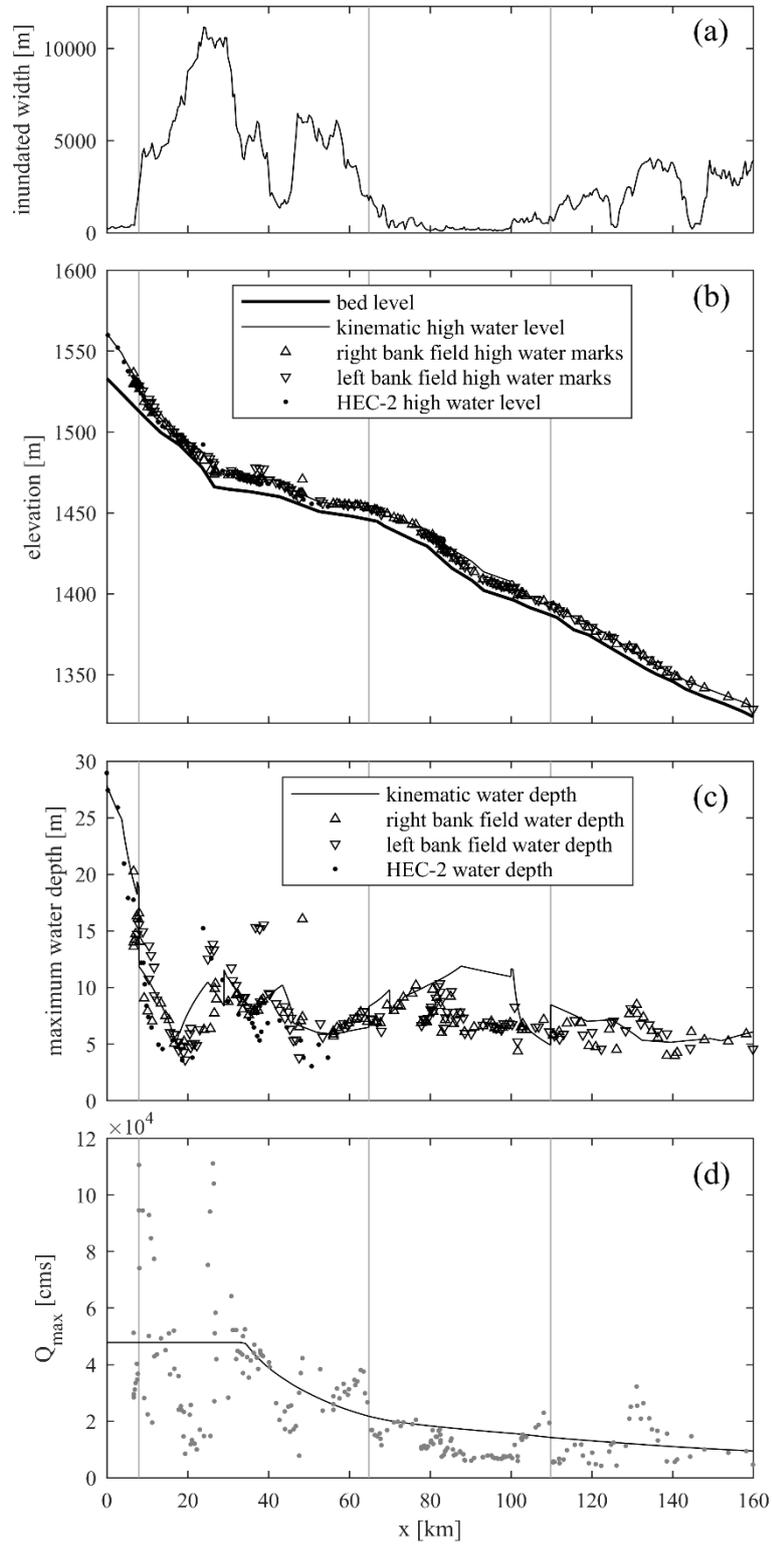

**Fig. 17.** Simulated and surveyed results for the Teton case. Gray lines: transitions between U and V shaped valleys. (a) Inundated widths along the valley; (b) bed elevation and maximum stage; (c) maximum depth along the valley; (d) maximum discharge attenuation calculated by the kinematic model (line) and back-calculated from the surveyed depths (dots). Four such values of $Q_{max}$ (1.7, 1.8, 2.2 and $3\times10^5$ cms) are beyond the axis range.

Using the assumed rating curves and measured water depths, we can also back-calculate the maximum flow discharge in the valley, and compare this signal with the discharge attenuation profile predicted by the kinematic simulation. Due to local departures of the flow from the assumed rating curve behavior, the back-calculated signal is rather noisy. The overall trend, however, whereby the maximum discharge gradually decreases with distance, is captured well by the kinematic model. In this case, the upstream hydrograph peaks sufficiently rapidly, and the post-flood survey covers a sufficiently long distance down valley to record a significant attenuation of the maximum discharge. Comparing Fig. 17a, b and 16d shows that attenuation is hastened in the area where the valley is wider and flatter. Conversely, discharge attenuation is delayed when the valley again becomes narrower and steeper. Overall, the proposed kinematic model is shown to capture well the main features of the flood behavior in this highly non-uniform valley.

### 5.3. Sudden failure of Malpasset Dam

Built across a narrow gorge of the Reyran River, France, the Malpasset arch dam failed catastrophically in 1959, releasing a 40-m high dam-break wave and causing multiple casualties. Because its width increases greatly from upstream to downstream, the valley is again highly non-uniform, as illustrated in Fig. 18. The data available, collected by Goutal (1999), include the valley topography from 1931, maximum water levels surveyed by the police after the flood, times of arrival of the flood front at three electric transformers, and the results from scaled physical model experiments conducted in 1964 by the Laboratoire National d'Hydraulique (LNH). The locations of the post-flood records, the three electric transformers, and the gauges used in the physical model are shown in Fig. 19 as white triangles, black triangles and white dots, respectively.

Since no discharge measurements are available, we use the model of Appendix B to simulate the dam failure hydrograph. We calculate the required reservoir storage curve from the digital topography, and assume a sudden dam-break. The resulting storage curve and predicted discharge hydrograph are shown in Fig. 19.

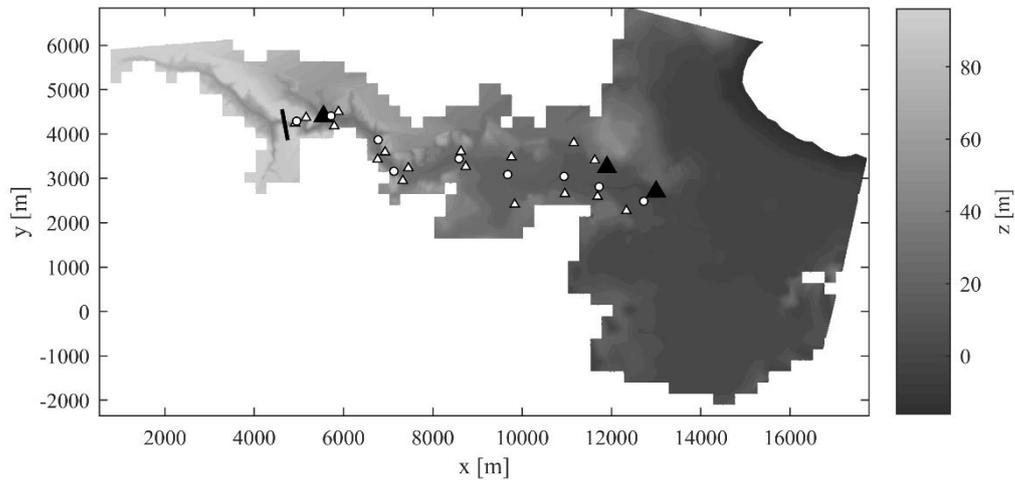

**Fig. 18.** Valley topography and location of the measurement points. Bold line: dam position; black triangles: electric transformers A, B, and C; white triangles: police survey points P1–P17; white dots: gauges S6-S14 used in the laboratory experiments. Data source: Goutal (1999).

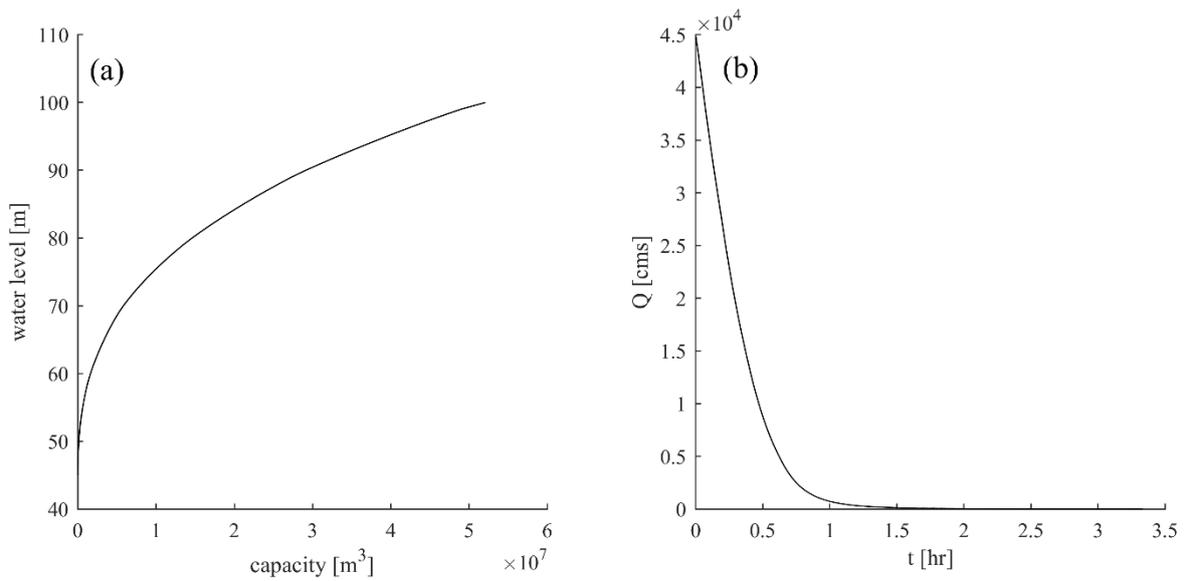

**Fig. 19.** Predicted dam failure hydrograph for the Malpasset case: (a) storage curve of Malpasset Reservoir, calculated from the digital topography; (b) simulated hydrograph.

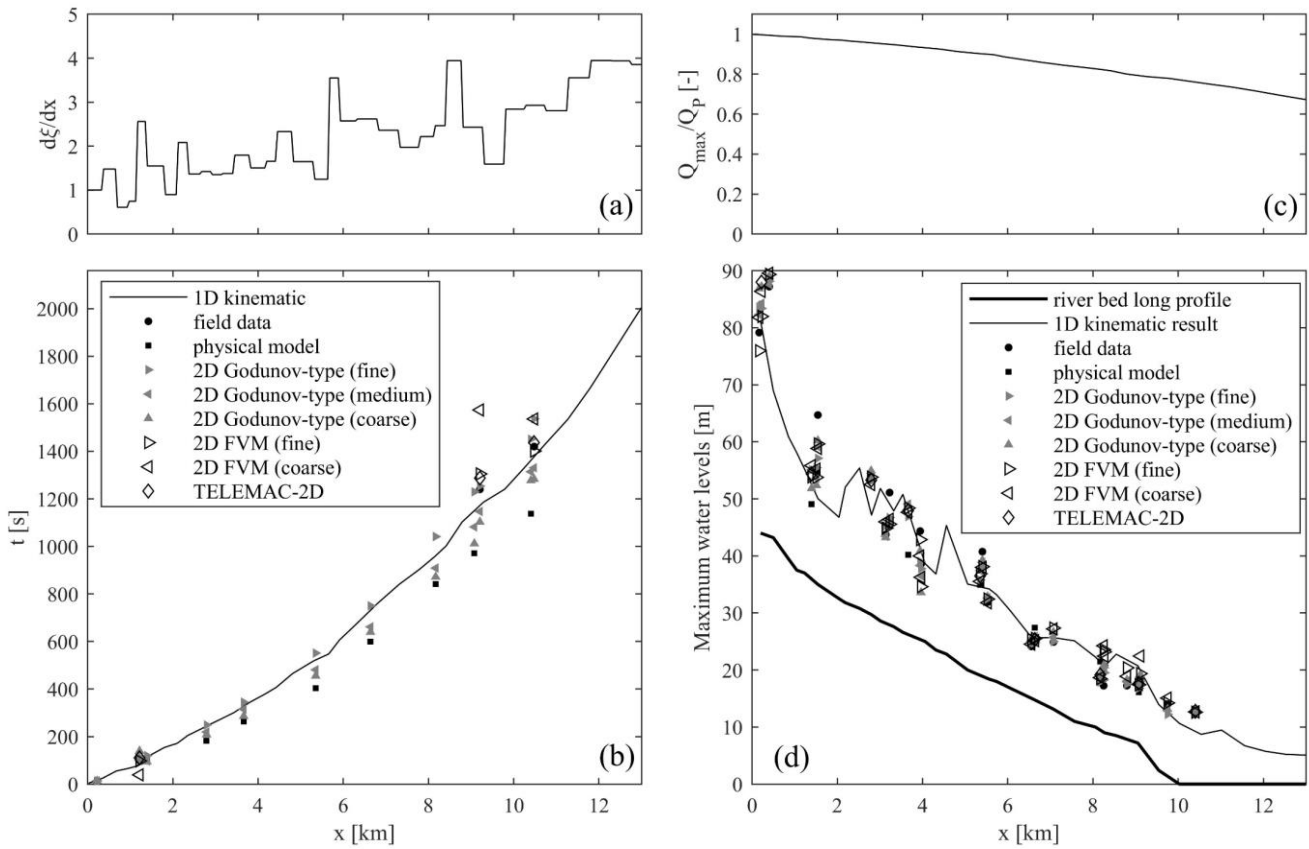

**Fig. 20.** Kinematic simulation results of the Malpasset dam break flood compared with the field, experiment, and 2D simulation data. (a) $d\xi/dx$ along the valley; (b) flood front propagation; (c) maximum discharge attenuation; (d) maximum water levels at the sampled cross-sections.

**Table 4.** RMSE between simulation results of various models and measurements.

| RMSE of | | Flood arrival times [s] | | Maximum water level [m] | |
|---|---|---|---|---|---|
| | | Electric transformer | Physical model | Field | Physical model |
| Kinematic wave | | 47.0 | 123.0 | 4.28 | 5.87 |
| 2D FVM (Valiani et al.) | Coarse | 207.2 | - | 3.70 | 3.27 |
| | Fine | 39.1 | - | 3.54 | 3.10 |
| TELEMAC-2D (Hervouet and Petitjean) | | 29.4 | - | - | 3.54 |
| 2D Godunov-type model (Kim et al.) | Coarse | 68.3 | 170.1 | 2.89 | 3.39 |
| | Medium | 74.3 | 65.1 | 2.75 | 3.86 |
| | Fine | 114.2 | 39.7 | 3.28 | 3.49 |

We calculated the rating curves of 39 evenly spaced cross sections using the measured valley cross sections with the Manning coefficient value $n = 0.033$ s/m$^{1/3}$ adopted by previous researchers (Valiani et al., 2002). We then approximate the valley by U-shaped cross sections and choose for the rating curve coefficients $m$ the values that best approximate the flow areas $A$ at discharge values $Q = 0.2Q_P$ and $0.8Q_P$. These simplified power law rating curves are used only to calculate the discharge profile evolution $Q(x,t)$ using the analytical solution. To calculate the resulting water levels with better accuracy, however, we use the more complex rating curves based on the original cross sections. A similar hybrid approach was earlier used by Capart (2013).

The analytical solution is again derived first in the $(\xi,t)$ plane, then converted to the $(x,t)$ plane using the $d\xi/dx$ profile calculated from the cross-section properties (Fig. 20a). In Fig. 20b, we compare the predicted wave front path (time of arrival) with the observations at the three transformers, and with corresponding arrival times calculated using the TELEMAC 2D software (Hervouet and Petitjean, 1999), a 2D dynamic FVM numerical model (Valiani et al., 2002), and a 2D Godunov-type model (Kim et al., 2014). In Table 4, we report the RMSE between results from various simulation method and the measurement obtained from field survey or physical model. The wave front curve predicted by the 1D kinematic model, which is influenced greatly by the $d\xi/dx$ profile, agrees well with the observed arrival times, and with those calculated using the more elaborate 2D models. According to Kim et al. (2014), the numerical error in 2D models is a function of mesh resolution and it may require a very fine resolution to minimize the numerical error. Nevertheless, the kinematic wave solution can provide the results with similar order of errors as the fine-resolution 2D model with very small computational cost.

In Fig. 20c, we plot the calculated maximum discharge as a function of distance. Although the valley is short (only 13 km), significant discharge attenuation occurs down valley because of the assumed instantaneous failure of the dam, producing a sudden rise of the hydrograph to its peak discharge. In Fig. 20d, finally, we compare the predicted maximum flood levels with those surveyed in the field, measured in the laboratory experiments, and computed using 2D models. To account for the influence of the valley topography, the predicted maximum flood levels are calculated using the raw rating curves instead of their power-law approximation. Using this hybrid approach, the error between the results of the kinematic model and the measured data are presented in Table 4. The differences between models are smaller than the differences between the field and measured data, and between the measurements acquired along the left and right sides of the valley. Overall, the maximum flood depth decreases from upstream to downstream, due both to the widening of the valley and to the gradual attenuation of the maximum discharge, itself accelerated when the valley becomes wider. These effects of valley non-uniformity are therefore found to be accurately captured by the proposed 1D kinematic model.

## 6. Conclusion

In this paper, we proposed an analytical approach to the kinematic routing of dam-break floods, in conditions more general than those addressed previously. The approach can handle generic upstream hydrographs, as well as non-uniform downstream valleys. Generic hydrographs are accommodated by applying the Gauss-Green theorem to characteristic bounded regions, yielding simple, compact formulas for the wave front path and maximum discharge attenuation. Results can then be extended to non-uniform valleys by simply rescaling the distance coordinate, taking into account variations in valley width, slope and friction factor. For special cases, the approach reproduces analytical solutions derived earlier for uniform valleys. Applied to idealized non-uniform valleys, the approach was also checked to agree with the results of numerical computations.

Most importantly, the approach is sufficiently general to be applied to actual field cases. For such cases, no two upstream hydrographs are the same, due to differences in dam failure mode and reservoir geometry. The downstream valley, moreover, is more likely than not to be highly non-uniform. Because it can deal with such complications, the proposed approach provides a useful tool for the routing of actual dam-break floods. This was demonstrated for three well-documented field cases: the 2008 breaching failure of the Tangjiashan landslide dam, the 1976 piping failure of Teton Dam, and the 1959 sudden failure of Malpasset Dam. For all three cases, simulation results were found to be in good agreement with field observations, and with results from more elaborate 2D dynamic models.

The proposed approach is not meant to compete with or replace the more elaborate models, but to provide a complementary tool with various advantages of its own. Operationally, the proposed kinematic model can simulate dam-break floods at minimal computational cost. This can be a significant advantage for the simulation of long valleys, multiple dam failure scenarios, or real-time forecasts. In addition, the model can be run subject to very limited data requirements. It does not require a detailed two-dimensional mesh or closely spaced cross-sections, and can make do with only a basic valley width and slope profile, or with a hydrograph reduced only to its peak discharge and time to peak. Conceptually, however, the model may be even more useful. Whereas elaborate models or field observations are often difficult to interpret, the proposed model greatly clarifies how dam-break flooding is affected by various contributing factors. Because the model is in analytical form, key outcomes like the time of arrival, maximum water level, and attenuation of the maximum discharge can be connected to specific properties of the upstream hydrograph and downstream valley.

Nevertheless, these advantages are subject to a number of important limitations. First, the proposed analytical solution is subject to various restrictions. As derived in this paper, it is only applicable to the simplest wave structure, featuring a single shock at the wave front. The proposed rescaling, moreover, is only applicable to separable rating curves, excluding cases like compound channels. Secondly, the one-dimensional kinematic wave equation provides only a first approximation to the behavior of dam-break waves. Over short distances, when backwater effects intervene, or in valleys with abrupt variations, non-local and inertial effects can be expected to become important. Thirdly, the proposed analytical solution is restricted to downstream topography that can be modeled as U shape or V shape valleys that gradually vary with distance from the dam. It is not applicable to more complex topography such as alluvial fans (e.g., Gallegos et al. 2009). Finally, whereas dam-break floods may cause significant geomorphic changes (see e.g. Capart et al. 2007), the approach is currently limited to fixed valley geometry. Addressing these limitations will require significant changes to the proposed approach, or the adoption of completely different methods.

**Appendix A: Numerical solution of the kinematic wave equation**

In this Appendix, we describe the computational scheme used in the paper to verify our analytical solutions for non-uniform valleys. The scheme selected is a first-order finite volume scheme, applicable to hyperbolic equations in the presence of shocks (Van Emelen, 2014; Lai and Khan, 2018). Accordingly, the continuity equation shown in Eq. (1) can be rewritten as

$$A_i^{k+1} = A_i^k + \frac{\Delta t}{\Delta x}\left(Q*_{i-1/2}^k - Q*_{i+1/2}^k\right),$$ (A.1)

where $\Delta x$ and $\Delta t$ are the special and temporal cell sizes, $A_i^k$ the flow area of $i$th cell at $k$th time step, and $Q*_{i+1/2}^k$ the average flux at the interface between $i$th cell and $i+1$th cell at $k$th time step. The flux $Q*_{i+1/2}^k$ can be solved by HLL solver which is developed by Harten, Lax and van Leer (1983). This solver has been widely used to solve shallow water problem in previous research, such as Caleffi et al. (2003), Valiani et al. (2002), Zhang et al. (2016), and Lai and Khan (2018). The flux $Q*_{i+1/2}^k$ can be written as

$$Q*_{i+1/2}^k = \frac{S_R Q_i^k - S_L Q_{i+1}^k + S_R S_L (A_{i+1}^k - A_i^k)}{S_R - S_L},$$ (A.2)

where $Q_i^k$ is the average discharge of $i$th cell at $k$th time step, and can be calculated by $A_i^k$ as shown below.

$$Q_i^k = m_i A_i^{k\,\alpha_i},$$ (A.3)

where $m_i$ and $\alpha_i$ are the coefficient and the exponent of the rating curve of $i$th cell. $S_L$ and $S_R$ represent the left and right wave speeds which can be estimated from

$$S_R = \max(\lambda_i, \lambda_{i+1}, 0), \quad S_L = \min(\lambda_i, \lambda_{i+1}, 0),$$ (A.4)

where $\lambda_i$ is the wave speed in the $i$th cell. Since wave speed in the kinematic solution is always positive (toward downstream), $S_L$ is a constant of 0 for every position. Therefore, the flux $Q*_{i+1/2}^k$ can be simplified into

$$Q*_{i+1/2}^k = Q_i^k.$$ (A.5)

It should be noticed that the backwater effect cannot be reflected in the proposed solution. Thus, the solution cannot be applied to cases in adverse slope reaches and upstream part of dams.

For the first flux $Q^{*k}_{-1/2}$, the upstream boundary condition (dam breach hydrograph) is adopted as

$$Q^{*k}_{-1/2} = Q_B(t^k). \tag{A.6}$$

In order to ensure the numerical stability of the explicit numerical solution, the time step $\Delta t$ have to conform with the Courant-Friedrichs-Lewy (CFT) condition (Harten et al., 1983).

$$Cr = \frac{\Delta t}{\Delta x} \lambda_{max} \leq 1, \tag{A.7}$$

where $Cr$ is the Courant number, and $\lambda_{max}$ is the maximum wave speed over the computational domain given by

$$\lambda_{max} = \max(\lambda_{i=0 \sim n}) = \max\left(\left.\frac{\partial Q_i}{\partial A_i}\right|_{Q_i^k}\right). \tag{A.8}$$

The above scheme is simple to apply to kinematic flood propagation problems in one-dimensional, non-uniform valleys. Because the scheme is only first order, however, a fine discretization is needed to obtain accurate results. Its computational cost is therefore much greater than the analytical method described in the paper.

**Appendix B: Dam failure hydrograph model**

In this Appendix, we describe and validate the model adopted to simulate dam failure hydrographs. Although the model is quite simple, it can account for various dam failure scenarios and general reservoir shapes. The model uses level-pool routing to calculate the evolving water volume in the reservoir, coupled with a dam erosion law governing the gradual failure of dams composed of erodible material (either natural or engineered).

The major variables in the coupled system are the two absolute levels: $z_L(t)$, the evolving lake water level, and $z_D(t)$, the time-varying level of the dam top. Two relative levels are defined by the relationships

$$\eta(t) = z_L(t) - z_D(t), \quad \delta(t) = z_D(0) - z_D(t), \tag{B.1}$$

where $\eta(t)$ is the relative depth between the lake water level and the dam top, and $\delta(t)$ the crest drop.

The level-pool routing equation (Henderson, 1966; Capart, 2013) can be written

$$A_L \frac{dz_L}{dt} = Q_U - Q_B, \tag{B.2}$$

where $A_L$ is the lake surface area depending on $z_L$, $Q_U$ the upstream discharge into the lake, and $Q_B(t)$ the time-varying outflow discharge through the breach. The breach is assumed to be a rectangular valley with width of $b_B$. Hence, the outflow discharge through the breach can be determined by using the broad-crested weir equation (Henderson, 1966)

$$Q_B = \sqrt{\frac{8}{27}} g^{1/2} b_B \eta^{3/2}. \tag{B.3}$$

The volumetric erosional rate of the dam is assumed to be governed by the stream power law (Lane, 1955; Capart et al., 2010, 2013)

$$\frac{d\forall_E}{dt} = -KQ_B S = KQ_B \frac{\partial z}{\partial x}, \tag{B.4}$$

where $K$ is a dimensionless transport coefficient, $-S(x,t) = \partial z / \partial x$ the local bottom inclination of the outflow valley. According to Capart (2013), the outflow valley bottom converges asymptotically towards the initial profile of the downstream dam face. Hence, $S(x,t)$ can be substituted by the constant $S_D$, the inclinations of downstream faces of the dam. The volume of material eroded from the dam is the product of the valley width, $b_B$ and the longitudinal section area of the breach, which is a function of $\delta(t)$

$$\forall_E = \lambda_B b_B \delta^{2\gamma}, \tag{B.5}$$

where $\lambda_B$ and $\gamma$ are constant coefficients dependent on dam geometry, material composition, and failure scenario.

We define

$$\alpha = \frac{1}{2}\left(\frac{2}{3}\right)^{3/2} \frac{KS_D g^{1/2}}{\lambda_B} = \frac{1}{2}\left(\frac{2}{3}\right)^{3/2} K_L S_D g^{1/2}, \quad \beta = \left(\frac{2}{3}\right)^{3/2} \frac{g^{1/2} b_B}{A_L},$$

(B.6)

where $K_L$ is equal to $K/\lambda_B$. The governing equations, Eq. (B.2)(B.4) then become the coupled system of ODEs

$$\frac{dz_L}{dt} = -\beta \eta^{3/2} + \frac{Q_U}{A_L},$$

(B.7a)

$$\frac{dz_D}{dt} = -\alpha \frac{\eta^{3/2}}{\delta^\gamma}.$$

(B.7b)

The system is an initial value problem which can be numerically solved with given initial conditions. Besides, it also needed to be take into account that the maximum breach depth $d_B = \max(\delta) = z_D(0) - z_D(\infty)$ is equal to $z_D(0) - z_{D,\min}$ when erosion reaches a non-erodible floor of which the level is $z_{D,\min}$.

Then the dam breach hydrograph can be obtained by substituting the solution of $\eta(t)$ into Eq. (B.3).

In order to solve the system, the parameters described in detail below need to be determined. $S_D$, $z_{D,\min}$ and $A_L(z_L)$ are usually known values in the cases of artificial dams, and easily obtainable values in the cases of naturally formed dams. $Q_U$ can be simply estimated from rainfall information. The breach width, $b_B$, can be estimated from the maximum breach depth, $d_B$. According to Singh and Scarlatos (1988), $b_B$ is usually smaller than 5 times of $d_B$ and has an average value of about 3 times of $d_B$. Once the abovementioned parameters have been determined, the solution is only dominated by $K_L$ and $\gamma$. As Fig. B.1 illustrates, the peak discharge $Q_P$ and the corresponding time to peak $T_P$ are impacted by $K_L$ and $\gamma$ in opposite directions. Here, the study uses $(d_B/6)^{3/2} g^{1/2} b_B$ and $\forall_B \left((d_B/6)^{3/2} g^{1/2} b_B\right)^{-1}/2$ to normalize $Q_P$ and $T_P$ for each case, where $\forall_B$ is the total drainage volume. The constant coefficient $K_L$ and $\gamma$ can be determined by searching the intersection of the contours of the normalized values of the estimated $Q_P$ and $T_P$. According to Capart (2013), if two dams break with the same scenario, their normalized dam failure hydrographs should be similar. Therefore, the estimated values of $Q_P$ and $T_P$ can be derived from the past dam failure hydrographs with the same dam break scenario.

In order to validate the numerical model, the study adopts three special cases with analytical solutions. The special cases include sudden dam breaks and gradual dam failures with ideal columnar reservoirs as well as sudden dam breaks with ideal triangular reservoirs, with subscript *sc*, *gc* and *st*, respectively. We assume there is no upstream inflow ($Q_U = 0$) is these cases. In the sudden dam break scenario (Hunt, 1984a), $z_D$ becomes $z_{D,\min}$ at the first moment of the dam failure process. Hence the coupled system of ODEs can be reduced into a single ODE (Eq. B.7a) and solved analytically. The resulted analytical hydrographs of sudden dam break scenario in both columnar and triangular reservoirs can be yielded as Eq. (29) and (52) with peak discharges

$$Q_{P,sc} = Q_{P,st} = \left(\frac{2}{3}d_B\right)^{3/2} g^{1/2} b_B \tag{B.8}$$

happen at $T_{P,sc} = T_{P,st} = 0$. That is to say, $Q_P = 8$ and $T_P = 0$ in both sudden dam break cases.

The analytical hydrograph of gradual dam failures with ideal columnar reservoirs is proposed by Capart (2013) as Eq. (31), with peak discharge

$$Q_{P,gc} = \left(\frac{1}{6}d_B\right)^{3/2} g^{1/2} d_B \tag{B.9}$$

happens at

$$T_{P,gc} = \frac{3A_L}{(Q_P g b_B^{\,2})^{1/3}}. \tag{B.10}$$

That is, $Q_P = 1$ and $T_P = 1$ in this case.

According to Fig. B.1, the sudden dam break scenarios in both reservoirs with $Q_P = 8$ and $T_P = 0$ are the extreme cases of the model which happen as $K_L$ and $\gamma$ are both large enough. Here the study calculates the numerical hydrographs with adopting $K_L = 0.2$, $\gamma = 2.4$ for each case. On the contrary, the gradual dam failure scenarios happen when $K_L$ and $\gamma$ are both small with columnar reservoir. Here we adopt $K_L = 0.0306$ and $\gamma = 1.0074$ for the numerical model by finding the intersection of the contours of $Q_P = 1$ and $T_P = 1$ in Fig. B.1b. In Fig. B.2, the numerical hydrographs (dash lines) match the analytical hydrographs (solid lines) well. Therefore, the numerical model is well validated and can be applied to a variety of reservoir shapes and dam failure scenarios.

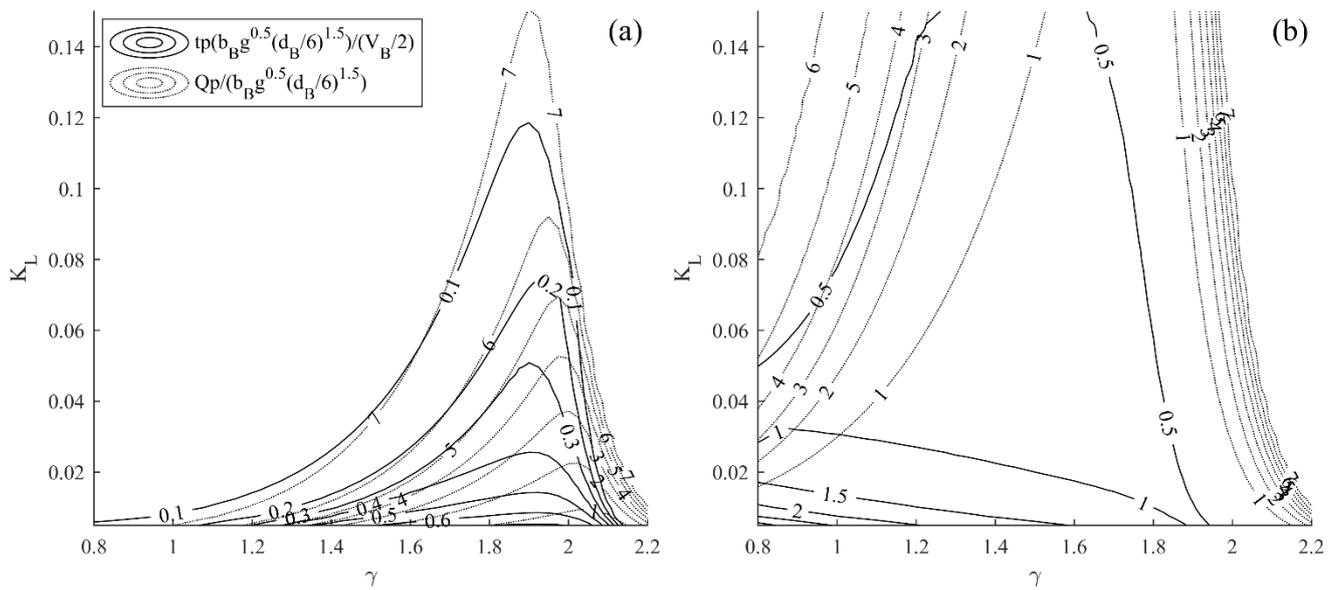

**Fig. B.1.** The influence of the coefficient $K_L$ and $\gamma$ on $Q_P$ and $T_P$. (a) Dam failure with no inflow triangular reservoirs; (b) dam failure with no inflow columnar reservoirs.

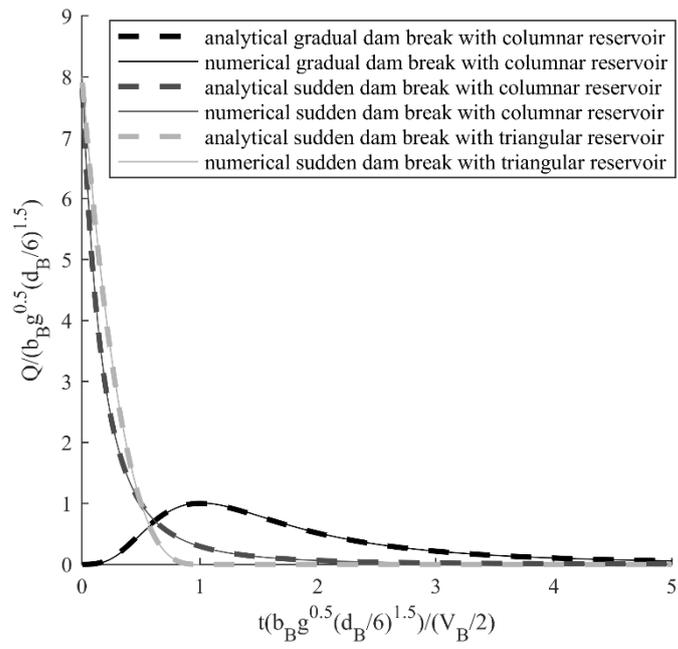

**Fig. B.2.** The comparison of the numerical hydrographs and the analytical hydrographs.